\documentclass[twoside,11pt]{article}

\usepackage{jmlr2e}
\usepackage{amsmath}
\usepackage{enumerate}
\usepackage{natbib}
\usepackage{url} 
\usepackage[linesnumbered,ruled,vlined]{algorithm2e}
\usepackage{verbatim}
\usepackage{mathrsfs}
\usepackage{nccbbb}
\usepackage{amssymb}
\usepackage{bm}
\usepackage{color}
\usepackage{multirow}
\usepackage{tabu}
\usepackage{ulem}
\usepackage{graphicx}
\usepackage{subcaption}
\usepackage{lineno}
\linenumbers
\usepackage{booktabs}
\usepackage{multirow}
\newcommand{\V}[1]{{\bm{\mathbf{#1}}}}
\newtheorem{condition}{Condition}

\newcommand{\vspacereduce}{\vspace{-0.5em}}

\definecolor{m_red}{rgb}{0.858, 0.188, 0.478}

\usepackage{lastpage}
\jmlrheading{24}{2023}{1-\pageref{LastPage}}{7/19; Revised
5/23}{5/23}{19-590}{Zhang, Song, Lu and Zhu}
\ShortHeadings{title}{Zhang, Song, Lu and Zhu}

\ShortHeadings{Distributed Community Detection in Large Networks}{}
\firstpageno{1}

\begin{document}
\nolinenumbers
\title{Distributed Community Detection in Large Networks}

 \author{\name Sheng Zhang \email szhang37@ncsu.edu \\
         \name Rui Song \email rsong@ncsu.edu \\
         \name Wenbin Lu \email wlu4@ncsu.edu\\
       \addr Department of Statistics\\
       North Carolina State University\\
       Raleigh, NC 27607, USA
       \AND
       \name Ji Zhu \email jizhu@umich.edu \\
       \addr Department of Statistics\\
       University of Michigan\\
       Ann Arbor, MI 48109, USA}

\editor{Edo Airoldi}

\maketitle

\begin{abstract}
    Community detection for large networks poses challenges due to the high computational cost as well as heterogeneous community structures.
    In this paper, we consider widely existing real-world networks with ``grouped communities'' (or ``the group structure''), where nodes within grouped communities are densely connected and nodes across grouped communities are relatively loosely connected.
    We propose a two-step community detection approach for such networks.
    Firstly, we leverage modularity optimization methods to partition the network into groups, where between-group connectivity is low. 
    Secondly, we employ the stochastic block model (SBM) or degree-corrected SBM (DCSBM) to further partition the groups into communities, allowing for varying levels of between-community connectivity.
    By incorporating this two-step structure, we introduce a novel divide-and-conquer algorithm that asymptotically recovers both the group structure and the community structure. 
    Numerical studies confirm that our approach significantly reduces computational costs while achieving competitive performance. 
    This framework provides a comprehensive solution for detecting community structures in networks with grouped communities, offering a valuable tool for various applications.

\end{abstract}


\section{Introduction}\label{sec:1}

For many real-world networks, such as social networks, airline route networks, and citation networks \citep{egghe1990introduction}, community structures are commonly observed as functional modules, i.e., nodes belonging to the same community share similar connectivity patterns.
Finding such community structures can be useful for exploring, modeling, and understanding networks \citep{rohe2011spectral}. 
The stochastic block model (SBM)  \citep{holland1983stochastic, snijders1997estimation}
has been widely used for modeling the community structure in networks and has shown appealing empirical and theoretical properties. 
Community detection based on the SBM is a nontrivial task, as optimizing the likelihood function over all possible community labels is an NP-hard problem \citep{bickel2009nonparametric}.
Various algorithms have been developed for community detection based on the SBM in the literature.  
For example, \cite{daudin2008mixture} considered the variational EM algorithm by optimizing the community assignment and probability matrix iteratively. 
\cite{rohe2011spectral} proposed the spectral clustering algorithm by analyzing the Laplacian of the adjacency matrix for high-dimensional SBM.
\cite{amini2013pseudo} proposed pseudo-likelihood based algorithms for network community detection tasks.
However, when dealing with very large networks, the computational cost is often too high and causes challenges.
In the variational EM method, each EM update requires $O(n^2)$ computations \citep{snijders1997estimation,handcock2007model}, where $n$ is the number of nodes in the network; 
for the spectral clustering method, the complexity for each iteration is $O(mK + nK^2)$, where $m$ is the number of edges and $K$ is the number of communities \citep{white2005spectral};
and the pseudo-likelihood algorithm lacks the guarantee of convergence \citep{wang2020fast}.
Besides, if the size of communities is unknown, selecting the number of communities also requires high computational cost. 
For example, \cite{wang2017likelihood} showed that directly using likelihood-based model selection methods requires an exponential number of summands and would become intractable as $n$ grows. 
Even with the variational likelihood EM update algorithm, each step will require $O(n^2)$ computations.
Note that these methods all work and only work with the entire network.
To overcome the computational burden for community detection in large networks, a natural idea is to first divide the large network into several sub-networks, and then communities are detected within each sub-network in a parallel way. 
The random division would break the community structure and lead to erroneous community detection results.
Therefore, a key question is how to divide the large network into sub-networks in a sensible way.

In many real-world applications, it has often been observed that communities in large networks can be divided into groups such that nodes (or communities) within the same group have relatively high link probabilities, while nodes in different groups have much lower link probabilities. 
One such example is the airline route network\footnote{https://openflights.org/data.html} which contains routes among airports spanning the globe.
It can be seen that the group structure in the airline route network corresponds to geographical regions, where connections within the same region are much denser than connections across regions.
Within each region (group), the airports are further divided into communities depending on connections due to different levels of economies, areas or politics. 
See Section \ref{subsec:6.1} for more details.

Another illustrative example is the Facebook ego network\footnote{https://snap.stanford.edu/data}, which is created based on the ``friend lists'' of 10 ego people (see Figure \ref{fig:egofb} in Appendix).
Naturally, people connected to the same ego person have higher link probabilities among themselves than to people connected to different egos. Therefore, it is reasonable to assume that there are 10 groups corresponding to the 10 egos in the Facebook ego network. Within each group, people could belong to different friend circles (communities), such as the colleague circle, the family circle, the classmate circle, etc, of the ego person. 
See Section \ref{subsec:6.2} for more details.

\begin{figure}[!htb]
\begin{subfigure}{0.5\textwidth}
\centering
\vspace{1cm}
\includegraphics[width=0.9\linewidth]{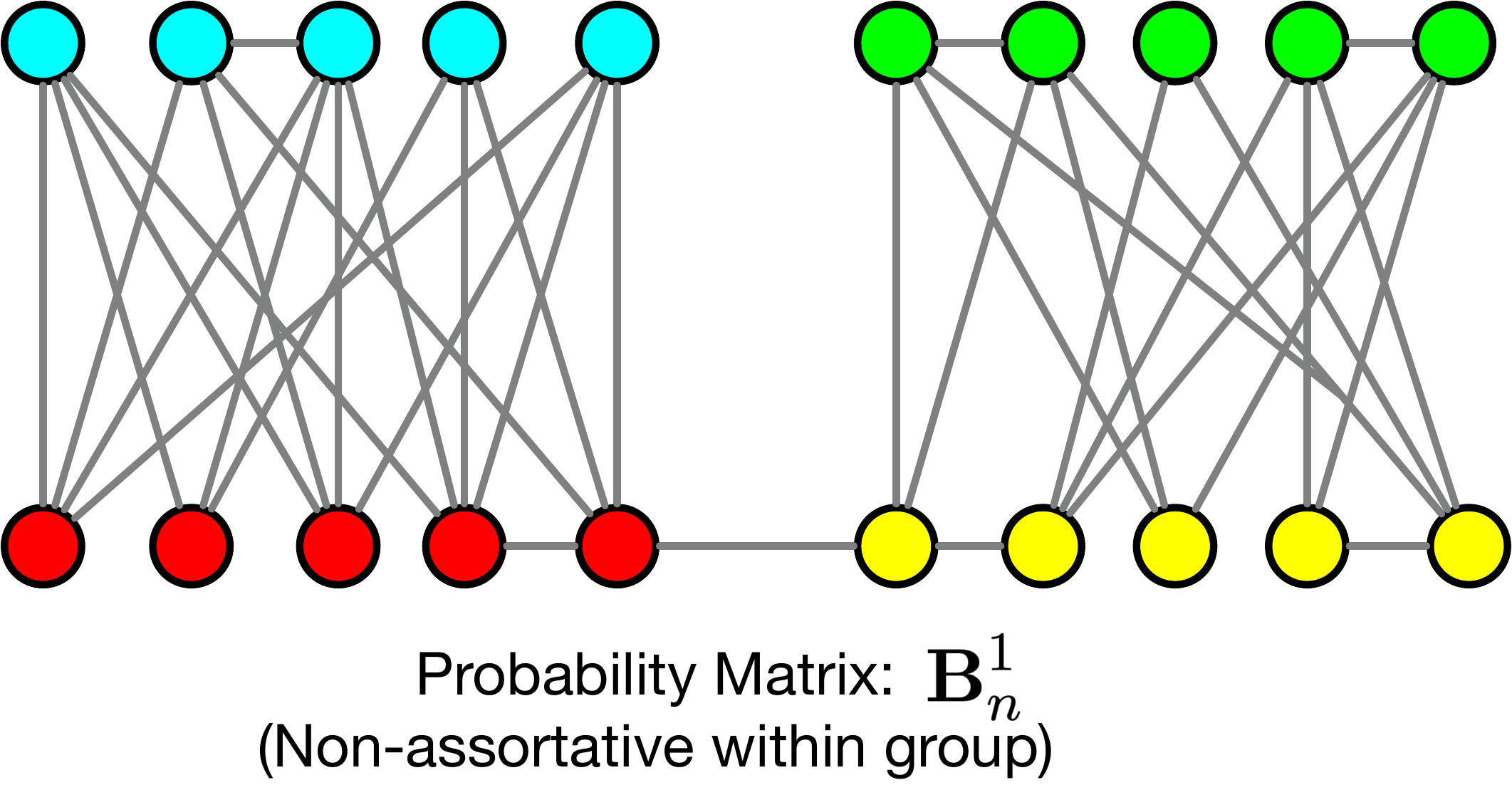}
\caption{}
\end{subfigure}\hfill
\begin{subfigure}{0.5\textwidth}
\centering
\includegraphics[width=0.9\linewidth]{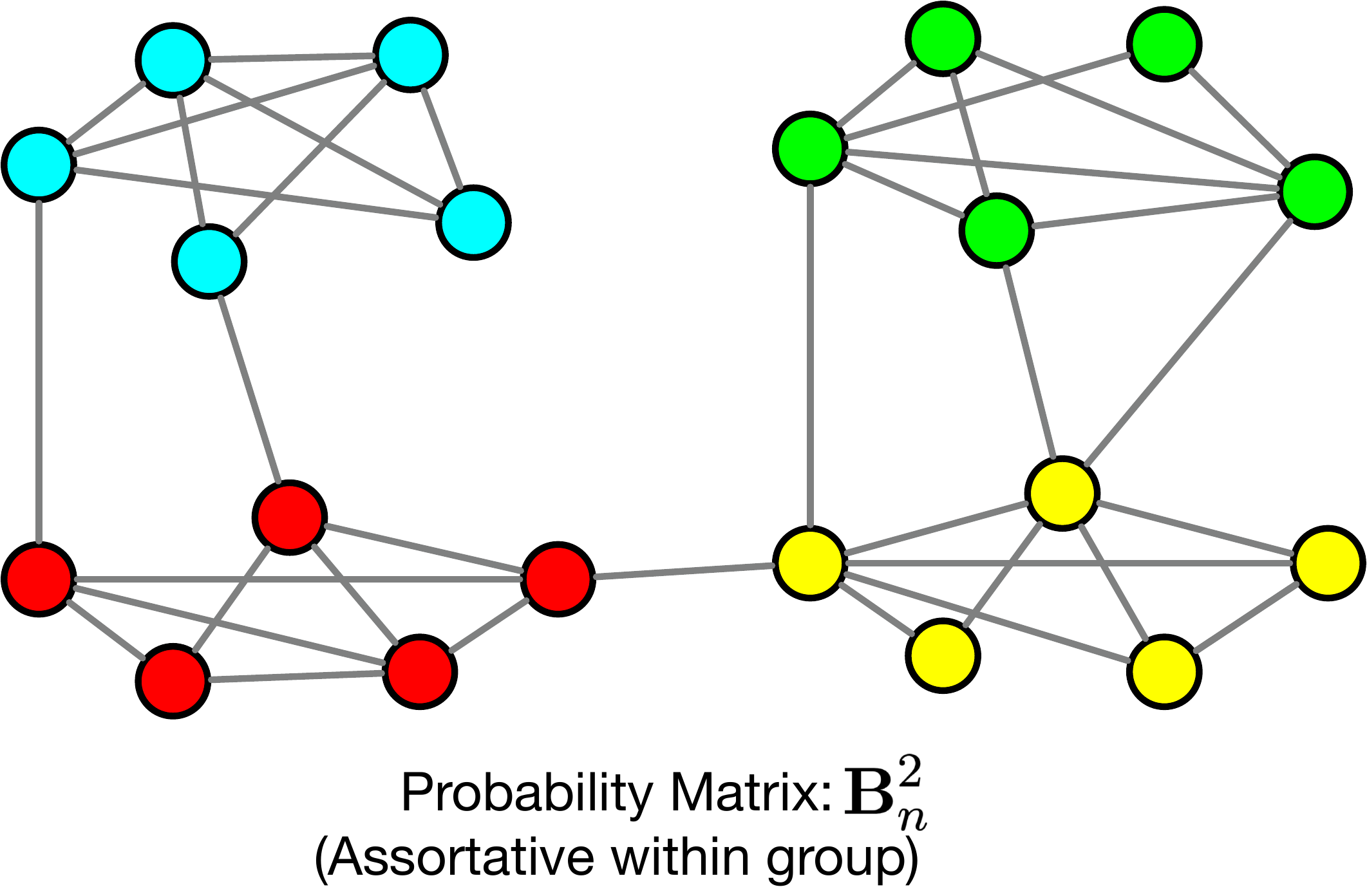}
\caption{}
\end{subfigure}
\caption{An illustration for the grouped community structure: the yellow and the green communities belong to the same group; the red and blue communities are in another group. In panel (a), communities within the same group are non-assortative (links are dense across communities); and in panel (b), communities within the same group are assortative (links are dense within communities). 
}\label{fig:illustration}
\end{figure}

In this paper, we consider large networks with a ``grouped community structure''. An illustration of the ``grouped community structure'' is shown in Figure \ref{fig:illustration}(a) and (b), where the network consists of four communities, with the green and the yellow communities belonging to one group and the red and blue communities belonging to the other group. The groups are such that nodes (or communities) within the same group have much denser links than those between different groups. 
In Figure \ref{fig:illustration}(a), communities within the same group are non-assortative, while Figure \ref{fig:illustration}(b) shows an example that communities within the same group are assortative.
Our goal has two folds: divide the large network into sub-networks according to the group structure so that it does not break the community structure within each group during the division; consistently conduct community detection within each sub-network. 

Towards this goal, we adopt a division method by optimizing the modularity \citep{newman2004finding}, which is a widely used measure for the strength of division of a network into groups. 
Specifically, modularity is defined as the fraction of edges that fall within the divided groups minus the expected fraction if edges were distributed at random \citep{newman2012communities}.
The larger the modularity, the divided groups will have denser connections within the groups but sparser connections between the groups. 
Many modularity optimization algorithms, such as the edge-betweenness method \citep{newman2004finding}, the fast-greedy method \citep{blondel2008fast} and the exhaustive modularity optimization via simulated annealing method \citep{guimera2004modularity}, have been developed to divide a network into groups.
These methods are usually model-free and can be implemented efficiently for large networks.
For example, the computational cost for the fast greedy algorithm is $O(n\hbox{log}^2 n)$, which is nearly linear in the size $n$ \citep{fortunato2016community}.

Although the modularity optimization methods are computationally efficient, they cannot be used to recover the refined community structure within groups since communities within the same group may have relatively more non-assortative links. 
This motivates us to adopt a novel divide-and-conquer scheme for distributed community detection for large networks. 
Specifically, we propose to first divide a large network into groups using the modularity-based method and then conduct community detection using the model-based method within each group. 
We show that our proposed divide-and-conquer method can estimate the community labels consistently for networks with the group structure under certain conditions.

The remainder of the paper is organized as follows. In Section \ref{sec:2}, we describe in detail the SBM with the grouped community structure. In Section \ref{sec:3}, we present the proposed distributed community detection algorithm. Theoretical properties of the proposed distributed community detection algorithm and extension to the Degree-Corrected SBM (DCSBM) are given in Section \ref{sec:4}. In Section \ref{sec:5}, we conduct simulation studies to examine the performance of the proposed method and compare it with other community detection methods. In Section \ref{sec:6}, we apply the proposed method to an airline route network and a Facebook ego network.
We conclude the paper with discussions in Section \ref{sec:7}. All technical proofs are provided in the Appendix. 
\section{\label{sec:2} SBM with Grouped Community Structures}

Consider an undirected and unweighted network $\V N = (\V{V},\V A)$, where $\V V = \{v_1,v_2,...,v_n\}$ is the set of nodes and $\V A = (A_{ij})$ is the adjacency matrix with $A_{ij} = 1$ representing a link between nodes $v_i$ and $v_j$, and 0 otherwise. 
First, we assume that the node set $\{v_1,v_2,...,v_n\}$ can be partitioned into $G$ disjoint groups $\mathcal{G}_1,\mathcal{G}_2,...,\mathcal{G}_G$. 
In addition, each group $\mathcal{G}_t$, is further partitioned into $K_t$ disjoint communities $\mathcal{C}_{t,1},\mathcal{C}_{t,2},...,\mathcal{C}_{t,K_t}$.
Second, the link probabilities are denoted by the matrix $\V B_n \in \mathbb{R}^{K \times K}$, where $K = \sum_{t=1}^GK_t$ is the total number of communities in the network. It is assumed that for any two nodes $v_i \in \mathcal{C}_{t,a}$ and $v_j \in \mathcal{C}_{s,b}$, they are connected by an edge with the probability $B_{n,\sum_{i = 0}^{t - 1}K_i + a, \sum_{j = 0}^{s - 1}K_j + b}$, where $B_{n,h,l}$ is the $(h,l)$th element in $\V B_n$ and $K_0$ is set as $0$.
Hence, the connection probability between nodes only depends on their community labels. 
For group structure $\V g = (g_1, g_2, ..., g_n)^T$ where $g_i \in \{1,2,..,G\}$ denotes the group label for node $v_i$, $i=1,\dots,n$, we assume that the group labels $g_i$'s follow a multinomial distribution with parameters $\V \Pi = (\Pi_1,\Pi_2,...,\Pi_G)^T$.
For community structure, we define $\V c = (c_1, c_2, ..., c_n)$, where $c_i \in \{1,2,..,K\}$ denotes the community label for node $v_i$, $i=1,\dots,n$, and the community labels $c_i$'s follow a multinomial distribution with parameters $\V\pi = (\pi_1,\pi_2,...,\pi_K)^T$.  
With these notation, for a node $v_i \in \mathcal{C}_{t,a}$ (i.e. the $a$-th community in the $t$-th group), we have $c_i = \sum_{i = 0}^{t - 1}K_i + a$.
We also have $\Pi_t = \sum_{j=1}^{K_t}\pi_{\sum_{i = 0}^{t - 1}K_i + j}$, $t=1,...,G$, which indicates that the frequency of a specific group is the summation of the frequencies of the communities in that group.

Note that for the grouped community structure to be meaningful, one needs additional constraints on the group assignment.  Specifically, we introduce Condition \ref{cond:1} on the generative probability matrix $\V B_n$ such that links within groups are denser than links across groups.

\begin{condition}\label{cond:1}
For graph $\mathcal{G}$ with community label $\mathbf{c}$ and probability matrix $\mathbf{B}_n$.
Given the group assignment $\mathbf{g}$, the probability matrix $\V B_n$ satisfies that $B_{n,ab} > B_{n,0}$ for communities $a$ and $b$ that belong to the same group $g_{.}$, and $B_{n,ab} < B_{n,0}$ for communities $a$ and $b$ that belong to different groups, where $B_{n,0} = \sum_{ab}\pi_a\pi_bB_{n,ab}$ measures the averaged connection probability over all communities.
\end{condition}

Lemma \ref{lemma:0} shows that, under Condition \ref{cond:1}, the group assignment is well-defined and unique. Therefore, there is no identifiability issue regarding to group structure.

\begin{lemma}\label{lemma:0}
For generative SBM graph with probability matrix $\mathbf{B}_n$, if both group assignments $\V g^1$ and $\V g^2$ satisfy Condition \ref{cond:1}, then $\V g^1 = \V g^2$. 
\end{lemma}
\noindent\textbf{Proof:} Suppose $\V g^1 \ne \V g^2$, then there exist nodes $v_i, v_j$ such that $g^1_i = g^1_j$ and $g^2_i \ne g^2_j$. Under the Condition \ref{cond:1}, we have $B_{n,c_ic_j} > B_{n,0}$ and $B_{n,c_ic_j} < B_{n,0}$, which causes contradiction. Therefore, Lemma \ref{lemma:0} holds. \hfill\BlackBox

{To illustrate the SBM with the grouped community structure, we consider the toy examples in Figures \ref{fig:illustration}(a) and (b). Both networks are generated under the SBM with $n=20$, $K=4$, and $\V\pi = (1/4,1/4,1/4,1/4)^T $.
The link probability matrices are
$$\V B_n^1 = 
\begin{pmatrix} 
0.3 &  0.8&  0.01& 0.01 \\
0.8&  0.3&  0.02& 0.02 \\
0.01& 0.02& 0.3&  0.9 \\
0.01& 0.02& 0.9&  0.3
\end{pmatrix}  ~~\mbox{and}~~  \V B_n^2 = 
\begin{pmatrix} 
0.8&  0.3&  0.02& 0.02 \\
0.3&  0.7&  0.01& 0.01\\
0.02& 0.01& 0.9&  0.3 \\
0.02& 0.01& 0.3&  0.9
\end{pmatrix} $$
respectively. Note that both probability matrices satisfy Condition \ref{cond:1} with group size of two.  Therefore, both networks are SBM with the grouped community structure, i.e., in both networks, nodes between groups are loosely connected in comparison to nodes within the same group. In Figure \ref{fig:illustration}(a), within community nodes are relatively loosely connected while nodes across communities but within the group are relatively densely connected. While in Figure \ref{fig:illustration}(b), the within community nodes have dense links and nodes across communities have loose links.}

\section{\label{sec:3} A Distributed Community Detection Algorithm}

{In this section, we present a divide-and-conquer algorithm that first divides the entire network into disjoint groups and then identifies communities within each group in the SBM. The summarized algorithms are provided in Algorithm \ref{algo: D_CD}.}

\subsection{Group Division Based on Modularity Optimization} 

{Modularity, as defined by \citet{newman2006modularity}, is the fraction of the edges that fall within the given groups minus the expected fraction if edges were distributed at random. Hence the group structure can be recovered by maximizing the modularity}.

To evaluate the optimality of a group assignment $\V e$, we use the Erdos-Renyi modularity \citep{guimera2004modularity} defined as 
\begin{equation}
\label{eq:ERM}
Q_{ER}(\V e) = \sum_{i,j}(A_{ij} -L/n^2)I(e_i = e_j),
\end{equation}
where $L$ is sum of degrees as $L = \sum_{ij}A_{ij}$.
Note that modularity $Q_{ER}(\V e)$ ranges from -1 to 1, and a large value of modularity indicates good performance for the group assignment $\V e$, i.e. the group division based on $\V e$ produces dense links within groups and sparse links between groups. 

Exact modularity maximization is an NP-hard problem, which may be intractable for large networks. Henceforth, several approximate methods have been developed to find decent solutions that approximately maximize the modularity. In our implementation, we adopt the fast greedy algorithm \citep{clauset2004finding} to divide a large network into groups by optimizing the modularity, which is a widely used hierarchical clustering algorithm for group detection. The basic idea of the fast greedy algorithm is to start with each node being the single member of a group. Then we repeatedly concatenate the two groups whose amalgamation produces the largest increase in modularity. For a network of $n$ nodes, after $n-1$ such joins, a single group remains and the algorithm stops. The entire process can be represented as a tree whose leaves are the nodes of the original network and whose internal joints correspond to the joins.  This dendrogram represents a hierarchical decomposition of the network into groups at all levels. Therefore, a cutoff point of the dendrogram can be selected based on the modularity or a given number of groups. 

{Since the fast greedy algorithm generates a dendrogram, the number of groups $G$ can be determined by choosing the partition with the greatest modularity during the training process without extra computation \citep{newman2004finding}.}
{One potential risk when using the maximal modularity to determine the groups is the ``resolution limit'' issue \citep{fortunato2007resolution}, which means maximizing modularity would miss substructures of a network, i.e. wrongly treat several small groups as one large group. However, in our method, the  ``resolution limit'' issue can be corrected in the following community detection step. 
}

\begin{algorithm}
\KwIn{the network $\V N = (\V V, \V A)$}
\KwOut{the estimated community labels $\hat{\V c}$ and estimated probability matrix $\hat{\V B}_n$}
\textit{PART I: Group Division} \\
Apply the fast greedy algorithm to the network $\V N$ and obtain the dendrogram $H$ \\
\textbf{If} $G$ is not given: \\
\ Choose an increment threshold $\delta$; initialize the group size $G = 1$; initialize the current modularity value $Q = 0$\\
 \ \   \For{$j=2; \ j < n; \ j = j + 1$}{
    Calculate the modularity value $Q_{j}$ with group size $j$ in dendrogram $H$\\
    \textbf{if} $Q_{j} - Q > \delta$: update the current modularity value $Q \leftarrow Q_{j}$ and the group size $G \leftarrow j$\\
    \textbf{else}: break the loop, output $G$} 
Divide the network $\V N$ into $G$ groups $\{\mathcal{G}_t\}_{\{t= 1,\ldots,G\}}$ based on the dendrogram $H$ \\
\textit{PART II: Community Detection}  \\
Initialization: $t = 1$\\
\Repeat{$t = G$}{
Determine the number of communities $\hat{K}_t$ in each group $\mathcal{G}_t$ using LRBIC; estimate the community labels $\hat{\V c}_t^{sub}$ in $\mathcal{G}_t$ using a SBM community detection method (such as VSBM or SSP) \\
\nl $t = t + 1$;\\
}
\textit{PART III: Combination}  \\
Obtain the community labels $\hat{\V c}$ by concatenating $\hat{\V c}^{sub}_t$, for $t= 1,\ldots,G$, and the estimated probability matrix $\hat{\V B}_n$ given $\hat{\V c}$.
\caption{Distributed Community Detection}
\label{algo: D_CD}
\end{algorithm}
\vspacereduce

\subsection{Distributed Community Detection Algorithm} \label{subsec: community_detection_step}
After dividing the network into disjoint groups, we conduct community detection within each group. Various methods can be adapted to estimate the community labels within each group, such as the variational EM algorithm based on SBM (VSBM) \citep{daudin2008mixture} and the regularized spherical spectral clustering (SSP) algorithm \citep{qin2013regularized}. The number of communities $K_t$ in group $t$ can be determined by the asymptotic likelihood ratio Bayesian information criterion (LRBIC) \citep{wang2017likelihood} which is a likelihood-based model selection method. We use $\V c^{sub}_t$ to denote the community labels in group $t$, where $\V c^{sub}_t = (c^{sub}_{t,1},c^{sub}_{t,2},\ldots,c^{sub}_{t,|\mathcal{G}_t|})$ and $|\cdot|$ denotes the cardinality of the set.  Let $\hat{\V c}^{sub}_t$ denote the estimated community labels in group $t$, for $t= 1,\ldots,G$. Then, the estimated community labels $\hat{\V c}$ of the entire network can be constructed by {concatenating detected community labels from all groups} as follows: for node $v_i$ in group $t$, if $\hat{\V c}^{sub}_t$ assigns $v_i$ to the $k$-th community in group $t$, then we denote the $i$-th element of $\hat{\V c}$ as $\sum_{j = 0}^{t - 1}\hat{K}_j + k$, where $\hat{K}_j$ is the estimated number of communities in group $j$ obtained from the previous step. Given the estimated community labels $\hat{\V c}$, we can further estimate the probability matrix $\hat{\V B}_n$ by the maximum likelihood estimation \citep{karrer2011stochastic}.

\subsection{Computational Cost for Proposed Method} 

{In terms of the computational cost, the division step takes $O(n\hbox{log}^2 n)$ computations in the fast greedy modularity method. Suppose group $\mathcal{G}_t$ (for $t=1,2,..,G$) contains $n_t$ nodes and $m_t$ edges. For the spectral clustering method, the computational complexity in $\mathcal{G}_t$ is $O(m_tK_t + n_tK_t^2)$. Define $K_{max} = \max(K_1,K_2,..,K_G)$ and $n_{max}, m_{max}$ in the same way. The computational cost for all groups is $O(G\times(m_{max}K_{max} + n_{max}K_{max}^2))$, which is much smaller than the original cost $O(mK + nK^2)$ as $n$ grows. For VSBM in the community-level detection step, the computational cost in each EM update is $O(G\times n_{max}^2)$ which is much smaller than applying VSBM to the whole graph. Furthermore, the computational time can be significantly decreased if distributed computing is used in the community detection step. It should also be noted that applying LRBIC to each group for selecting the number of communities has lower computational cost than to the whole network.} 

\section{\label{sec:4} Theoretical Properties}

In this section, we show that the proposed distributed community detection algorithm can consistently recover the community labels for networks with the grouped community structure. We state the main results in this section and provide all the proofs in the Appendix.



\subsection{Group Detection Consistency for the SBM}

We first formally formulate the group detection problem for networks with the grouped community structure.  Given the total number of groups $G$, we use $\V e$ to denote a group assignment $\V e = \{e_1,e_2,\ldots,e_n\}$, where $e_i$ is the group label for node $v_i$ and it takes value in $\{1,2,\ldots,G\}$. Define $\hat{\V g} = \mbox{argmax}_{\V e}{Q_{ER}(\V e)}$ as the estimated group labels. Our first goal is to show that $\hat {\V g}$ can consistently estimate the true group labels $\V g$. 

To establish group consistency for the proposed distributed community detection algorithm, we make the following assumptions for the probability matrix $\V{B}_n$ and the true group assignment $\V g$.  First, we note the probability matrix $\V{B}_n$ is not fixed, as otherwise, the expected degree for each node will grow proportionally to $n$ as $n \rightarrow \infty$, which is not realistic in practical applications. Specifically, we assume the following condition for the probability matrix $\V{B_{n}}$.

\begin{condition}\label{cond:2}
The probability matrix $\V{B}_n$ can be written as $\V B_{n} = \rho_n \V B$, where $\rho_n = \V{\pi}^T\V B_{n}\V{\pi} =  \sum_{ab}\pi_a\pi_bB_{n,ab}$ is the averaged probability of two nodes being linked. We assume $\rho_n \rightarrow 0$ as $n \rightarrow \infty$. 
\end{condition}

Condition \ref{cond:2} has been widely used in the literature for studying the consistency of community detection in networks \citep[e.g.][]{bickel2009nonparametric, lei2015consistency,  jin2015fast, zhao2012consistency, Abbe2017CommunityDA}.  Under Condition \ref{cond:2}, the expected degree of a node is given by $\lambda_n = n\rho_n$ and the expected total degree in the networks can be written as $u_n = n\lambda_n = n^2\rho_n$. If $n\rho_n = O(1)$, then the expected degree of a node is bounded as the size of the network grows to infinity. 

The strong and weak consistency of estimated group labels $\hat{\V g}$ can be defined by
$$ P(\hat{\V g} = \V g) \rightarrow 1 \ \ \ as \ n \rightarrow \infty, $$
and for any $\varepsilon > 0$,
$$ P\left[\left\{ \frac{1}{n}\sum_i 1(\hat{g}_i \ne g_i) \right\} < \varepsilon\right] \rightarrow 1 \ \ \ as \ n \rightarrow \infty, $$ 
respectively. In the next theorem, we establish the consistency for group detection. Note that all the results are up to permutation of the group labels.

\begin{theorem} \label{theo:1} Under the SBM, suppose $\V{B}_n$ satisfies Conditions \ref{cond:1} and \ref{cond:2}, then the estimated group labels $\hat{\V g}$ obtained by maximizing $Q_{ER}(\V e)$ are strongly consistent when $\lambda_n/\log(n) \rightarrow \infty$ and weakly consistent when $\lambda_n \rightarrow \infty$. 
\end{theorem}

{Theorem \ref{theo:1} shows that by maximizing the modularity and under certain conditions, the estimated group labels $\hat{\V g}$ can consistently recover the group structure in the SBM, without breaking the community structures within the groups.}

{In comparison to previous works, the key difference lies in the requirement on the link probability matrix in Condition \ref{cond:1}. For example, Theorem 2 in \cite{zhao2012consistency} requires the link probabilities within communities to be higher than a threshold while link probabilities across communities are lower than it, i.e. to achieve consistency, the community structure is required to be assortative. In our theory, the community-level assortative condition is relaxed to the group-level assortative condition, which leads to the divide-and-conquer algorithm (i.e. Algorithm \ref{algo: D_CD}) that works not only for assortative networks but also for non-assortative networks.}

{There are two key and unique challenges in the proof of Theorem \ref{theo:1}, which are not shared by previous works.  The first one is how to make sure the maximization of modularity maintains the community structure.  Motivated by the proof in \cite{zhao2012consistency}, we show that if the maximizer of the modularity's expectation can be denoted as a generic $G \times K$ matrix with at most one nonzero element in each column, then the corresponding group estimation is consistent. The second challenge is to show that Conditions \ref{cond:1} and \ref{cond:2} are sufficient to lead to the generic matrix with at most one nonzero element in each column.  The detailed proof is provided in the Appendix.}



\subsection{{Extension to the Degree-corrected SBM }}\label{dcsmb}

To allow for more flexible degree heterogeneity, we consider the degree-corrected stochastic block model (DCSBM) \citep{karrer2011stochastic}, an extension of the SBM. Out of simplicity, we use the same notations as in \ref{sec:2} in DCSBM. In addition to the parameters of the SBM, each node is also associated with a degree parameter $\theta_i$. For identifiability, we assume the expectation of the degree variable is 1, i.e. $E(\theta_i) = 1$. The edge $A_{ij}$ is a Poisson distributed random variable with $P(A_{ij} = 1 | \V c, \V \theta) = \theta_i\theta_j B_{n,c_ic_j}$, where $\V\theta = (\theta_1,\theta_2,..,\theta_n)$.

{The group structure in DCSBM is defined under the same idea as in grouped SBM: links within groups are denser than links between groups. Since the DCSBM is a generative model, the group structure can be defined by adding constraints on the probability matrix $\V B_n$ and the degree variable $\V \theta$ as in Condition \ref{cond:1dc}. Further, Lemma \ref{lemma:0dc} shows that the group structure in the DCSBM is well-defined.}

\begin{condition}\label{cond:1dc}
Suppose $\theta_i$ is a discrete random variable and takes value in the set $\{x_1, x_2, .., x_M\}$.  Let $\tau_{au}$ denote the frequency of community $a$ and the degree value $x_u$ (i.e. $\tau_{au} = P(c_i = a, \theta_i = x_u)$). Further, let $\tilde{\pi}_a = \sum_u\tau_{au}$ and $\tilde{B}_{n, 0} = \sum_{ab} \tilde\pi_a \tilde\pi_b B_{n, ab}$.  For the normalized matrix $\V \Gamma = \tilde{\V W} - (\tilde{\V W}\mathbf{1})(\tilde{\V W}\mathbf{1})^T$ where $\tilde{W}_{ab} = \tilde{\pi}_a\tilde{\pi}_bB_{n, ab} / \tilde{B}_{n,0}$, it satisfies that $ \Gamma_{n,ab} > 0$ for communities $a$ and $b$ that belong to the same group, and $\Gamma_{n,ab} < 0$ for $a$ and $b$ that belong to different groups.
\end{condition}

\begin{lemma}\label{lemma:0dc}
For generative DCSBM graph with probability $\mathbf{B}_n$, if both $\V g^1$ and $\V g^2$ satisfy Condition \ref{cond:1dc},  then we have $\V g^1 = \V g^2$. 
\end{lemma}

{Following the same idea as in Section \ref{sec:3} for the SBM, we can build the distributed community detection algorithm for the DCSBM. In the group division part, the modularity under the DCSBM is defined in Eq. \eqref{eq:DCERM}, which is a modification of the Erdos-Renyi modularity in Eq. \eqref{eq:ERM} by taking into account the degree heterogeneity, }
\begin{equation} 
\label{eq:DCERM}
Q_{DC}(\V e) = \sum_{i,j}(A_{ij} -d_id_j/L)I(e_i = e_j),
\end{equation}
{where $d_i = \sum_j A_{ij}$ is the degree of node $v_i$, and $\V e$, $\V O(\V e)$, $L$ are all defined the same as before. }
{As discussed in Section 3.1, exact modularity maximization is infeasible. Therefore, we again apply a fast greedy method to the degree-corrected modularity. Just like in the SBM, the number of groups can be obtained during the training process of the fast greedy method.}

In the community detection part, we first use LRBIC to determine the number of communities $K_t$, then we apply community detection methods such as the variational DCSBM method, spectral clustering or SCORE \citep{jin2015fast}, to identify the community label $\V c_t^{sub}$ within each group $t, t = 1,2,..,G$. The combination of the community detection part and the group detection part is the same as in Algorithm \ref{algo: D_CD}. 

Lastly, we show the group consistency result under the DCSBM in Theorem \ref{theo:1dc}. Based on the Theorem, the group detection part of the distributed community detection algorithm can consistently recover the group labels for the DCSBM with group structure.

\begin{theorem} \label{theo:1dc} Under the DCSBM, if the model satisfies Conditions \ref{cond:2} and \ref{cond:1dc}, then the estimated group labels $\hat{\V g}$ obtained by maximizing $Q_{DC}(\V e)$ are strongly consistent when $\lambda_n/\log(n) \rightarrow \infty$ and weakly consistent when $\lambda_n \rightarrow \infty$.
\end{theorem}

\subsection{Community Detection Consistency}


In this subsection, we establish the strong and weak consistency for the estimated community labels $\hat{\V c}$ obtained using the proposed distributed community detection algorithm. The definitions of the strong and weak consistency for the estimated community labels $\hat{\V c}$ are similar to those for $\hat{\V g}$: The strong and weak consistency of estimated community labels $\hat{\V c}$ can be defined by
$$ P(\hat{\V c} = \V c) \rightarrow 1 \ \ \ as \ n \rightarrow \infty, $$
and for any $\varepsilon > 0$,
$$ P\left[\left\{ \frac{1}{n}\sum_i 1(\hat{c}_i \ne c_i) \right\} < \varepsilon\right] \rightarrow 1 \ \ \ as \ n \rightarrow \infty, $$ 
respectively. 

\begin{theorem}\label{theo:2}
(Consistency of the Distributed Community Detection Algorithm) Provided that the estimated group labels $\hat{\V g}$ are strongly consistent and the adopted community detection algorithm is strongly (weakly) consistent, then the estimated community labels $\hat{\V c}$ are strongly (weakly) consistent.
\end{theorem}

\noindent\textbf{Remark:} {Note that Theorem \ref{theo:2} covers both the SBM and the DCSBM.}  If the number of groups $G$ is 1, Theorem \ref{theo:2} reduces to the community detection results established in \cite{bickel2009nonparametric} {for the SBM and that in \cite{zhao2011community} for the DCSBM}. 

Because the community level detection methods, including the SSP \citep{von2008consistency} and VSBM \citep{daudin2008mixture} {for the SBM and SCORE \citep{jin2015fast} for the DCSBM}, are consistent, our proposed distributed community detection algorithm is consistent provided that the estimated group labels are strongly consistent as stated in Theorem \ref{theo:2}.

\section{\label{sec:5}Simulation Studies}

In this section, we conduct simulation studies to evaluate the performance of the proposed distributed community algorithm.

\subsection{Comparative Methods and Implementation Detail}
5In SBM, we consider three baselines. First, FG denotes the ER modularity optimization (Eq \eqref{eq:ERM}) based on fast-greedy algorithm.\footnote{ https://igraph.org/r/doc/cluster\_fast\_greedy.html} The estimated community size and community label can be obtained simultaneously from the dendrogram generated by fast-greedy algorithm.
Second, VSBM denotes the variational EM algorithm for maximizing SBM likelihood function.\footnote{https://rdrr.io/cran/mixer/} The size of community is selected by LRBIC method where Kmax is set as the true community size.\footnote{LRBIC function in randnet R package: https://cran.r-project.org/web/packages/randnet/randnet.pdf}
Third, SSP denotes the spherical spectral clustering algorithm,\footnote{reg.SSP  function in randnet R package: https://cran.r-project.org/web/packages/randnet/randnet.pdf} the size of community is also selected by LRBIC method.

As comparison, we study two distributed community detection methods summarized in Algorithm \ref{algo: D_CD}. D-VSBM and D-SSP represent group division via modularity optimization and community detection within each sub-network is conducted using VSBM and SSP methods, respectively.
The group size is obtained from the group division step and community sizes within sub-networks are estimated from LRBIC method as well.

In DCSBM, three baselines are compared. Firstly, FG-DC denotes the DC modularity optimization (Eq \eqref{eq:DCERM}) based on fast-greedy algorithm. 
Secondly, VDCSBM denotes the variational EM algorithm for maximizing DCSBM likelihood function. Lastly, SCORE denotes the spectral clustering on ratios-of-eigenvectors community detection method.\footnote{https://rdrr.io/cran/ScorePlus/man/SCORE.html} 
In both VDCSBM and SCORE method, the size of community is selected through LRBIC method. We compare the baselines with distributed method, D-DCSBM and D-SCORE, to evaluate the performance of proposed methods.

\subsection{\label{subsec:5.2}Numerical Results for SBM and DCSBM}



\begin{figure}[ht]
\centering
\includegraphics[width=1\textwidth]{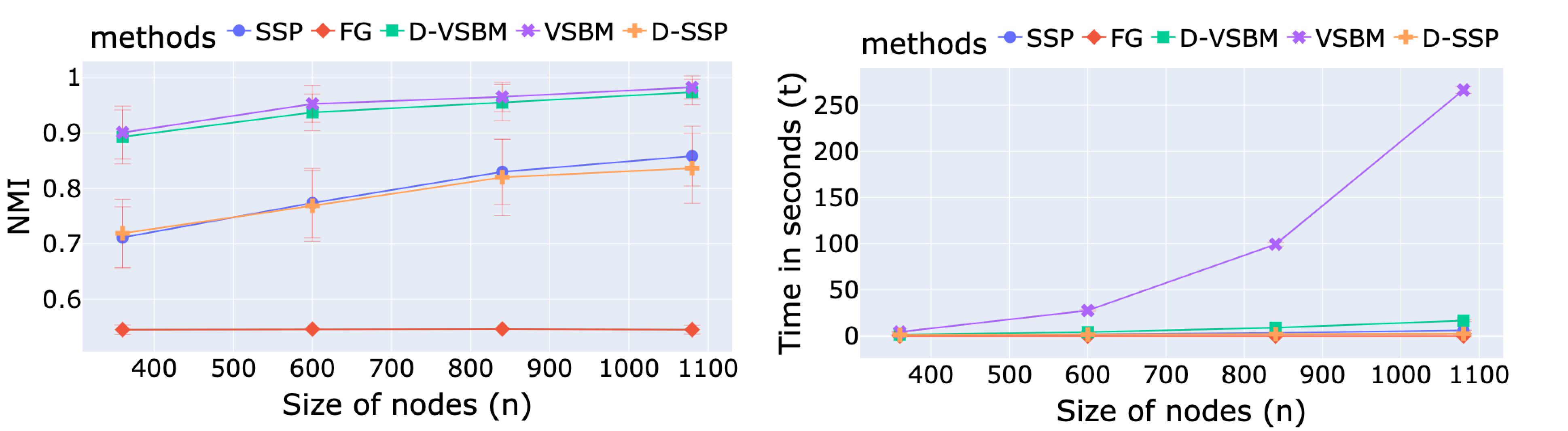}
\caption{Relatively small networks. The panels on the left show the NMI for all methods with  growing number of nodes $n$ under the SBM; the panels on the right show the corresponding computing times; the panel on the bottom is the group NMI.}
\label{Fig:comparemethods}
\end{figure}

\begin{figure}[ht]
\centering
\includegraphics[width=1\textwidth]{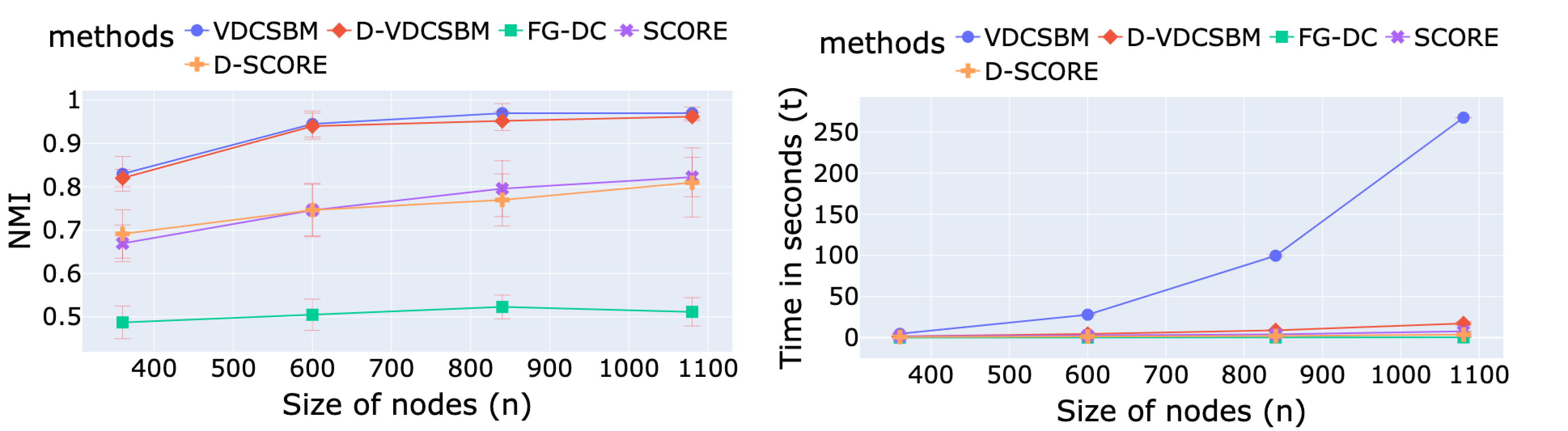}
\caption{The panels on the left show the NMI for all methods with growing number of nodes $n$ under the DCSBM. The panels on the right show the corresponding computing times; the panel on the bottom is the group NMI.}
\label{Fig:comparemethodsdcsbm}
\end{figure}

We start with grouped community structure SBM network with generation based on the description in Section \ref{sec:2}. Specifically, we generate networks using the SBM with $G = 4$ groups, the four groups contain 2, 3, 3 and 4 communities, respectively. The community frequencies are set as $\V \pi = (1/12,1/12,..,1/12)^T$. The probability matrix $\V B_n \in \mathbb{R}^{12 \times 12}$ is generated as follows: for nodes $v_i$ and $v_j$ belonging to the same group, $B_{c_ic_j} \sim \mbox{Unif}(1/100, 1)$; for nodes $v_i$ and $v_j$ from different groups, $B_{c_ic_j} \sim \mbox{Unif}(0, 1/100)$.
Hence, the generated $\V B_n$ satisfies Condition \ref{cond:1}. 
Therefore, the networks which are generated from $\V B_n$ have structure that community structure within a group is non-assortative while the groups remain as assortative. Under the setting of DCSBM, we follow the network generation strategy as in SBM, except that now we have the extra degree parameter $\V \theta = (\theta_1, \ldots, \theta_n)$, and they are generated independently as follows,
$$
P(\theta_i = mx) = P(\theta_i = x) = \frac{1}{2},
$$
where we set $m = 3/2$ and $x = 2/(m + 1)$ so that $E(\theta_i) = 1$ is satisfied as discussed in Section \ref{dcsmb}. For each setting, we repeat the simulation 100 times with various network sizes $n$.

To evaluate the performance of different methods of community detection, we use the normalized mutual information (NMI) index \citep{mori2003recognizing}  shown in Equation \eqref{eq:NMI} where $I(\cdot, \cdot)$ and $H(\cdot)$ are mutual information and entropy respectively. NMI ranges between 0 and 1, with larger values indicating better performance.  NMI equals 1 when the method recovers the true community structure perfectly. 

\begin{equation}\label{eq:NMI}
\begin{split}
    & NMI(\mathbf{c}, \mathbf{\hat{c}})  = \frac{2 \times I(\mathbf{c}, \mathbf{\hat{c}})}{H(\mathbf{c}), H(\mathbf{\hat{c}})} \\
    I(\mathbf{c}, \mathbf{\hat{c}}) & = \sum_{e_1 \in \mathbf{c}}\sum_{e_2 \in \mathbf{\hat{c}}} p_{(\mathbf{c}, \mathbf{\hat{c})}}(e_1, e_2) log(\frac{p_{(\mathbf{c}, \mathbf{\hat{c}})}(e_1, e_2)}{p_{\mathbf{c}}(e_1)p_{\mathbf{\hat{c}}}(e_2)}) \\
\end{split}
\end{equation}
We also compare the computational time for different methods to evaluate the scalability.
The results for both SBM and DCSBM are summarized in Figures \ref{Fig:comparemethods} and \ref{Fig:comparemethodsdcsbm}, respectively. 
We have three findings from the result. First, as the size of the network $n$ increases, except for the modularity-based method (FG and FG-DC), the NMI increases for all other methods in both SBM and DCSBM models. 
Also, we observe that likelihood-based method VSBM and VDCSBM are better than SSP and SCORE in NMI, respectively, while the computational time for VSBM and VDCSBM grows exponentially which means they are not scalable. 
Third, we find that the distributed method and non-distributed method are close in community detection performance, while distributed methods are scalable in computation time.


To study the performance in relatively large size of networks, we consider relatively large networks where the size of networks varies from 5e3 to 5e5 and the the size group $G$ also grows as the size of nodes increases.
Within each group, the community size is 5 with equal community frequency.
The probability matrix is designed as the small scale experiment above.
we summarize the result in Table \ref{Tab:comparemethodslargen}.  Computational time including community size selection using LRBIC and community label detection step, hence SSP and SCORE cannot be calculated when size of n is too large. 
We use G-NMI to denote the group level NMI which is defined as G-NMI=  NMI$(\mathbf{g}, \mathbf{\hat{g}})$.
From the result, we show the proposed distributed based methods works well even in large size of networks setting.
Also, the G-NMIs are large which indicate the proposed distributed method can maintain the group level structure well. 
The distributed based methods are also scalable compared to baseline methods.

\begin{table}[ht]
\begin{tabular}{@{}lllll|llll@{}}
\toprule
\multicolumn{5}{l|}{SBM}                                                                                                                                                                                           & \multicolumn{4}{l}{DCSBM}                                                                                                           \\ \midrule
(n, G, K)                                                                       & Method & NMI                                                     & t (s)                                          & G-NMI & Method   & NMI                                                      & t (s)                                              & G-NMI \\ \midrule
\multirow{4}{*}{\begin{tabular}[c]{@{}l@{}}(5e3, \\  1e1,\\  5e1)\end{tabular}} & VSBM   & \begin{tabular}[c]{@{}l@{}}0.97 \\ (0.005)\end{tabular} & \begin{tabular}[c]{@{}l@{}}6952\\ (20)\end{tabular} & -     & VDCSBM   & \begin{tabular}[c]{@{}l@{}}0.95 \\ (0.006)\end{tabular}  & \begin{tabular}[c]{@{}l@{}}7024\\ (20)\end{tabular} & -     \\
                                                                                & D-VSBM & \begin{tabular}[c]{@{}l@{}}0.97 \\ (0.008)\end{tabular} & \begin{tabular}[c]{@{}l@{}}42\\ (6)\end{tabular}    & 0.95  & D-VDCSBM & \begin{tabular}[c]{@{}l@{}}0.95 \\ (0.010)\end{tabular}  & \begin{tabular}[c]{@{}l@{}}48\\ (5)\end{tabular}    & 0.96  \\
                                                                                & SSP    & \begin{tabular}[c]{@{}l@{}}0.88 \\ (0.004)\end{tabular} & \begin{tabular}[c]{@{}l@{}}411\\ (8)\end{tabular}   & -     & SCORE    & \begin{tabular}[c]{@{}l@{}}0.90 \\ (0.007)\end{tabular}  & \begin{tabular}[c]{@{}l@{}}502\\ (9)\end{tabular}   & -     \\
                                                                                & D-SSP  & \begin{tabular}[c]{@{}l@{}}0.86 \\ (0.006)\end{tabular} & \begin{tabular}[c]{@{}l@{}}25\\ (4)\end{tabular}    & 0.95  & D-SCORE  & \begin{tabular}[c]{@{}l@{}}0.89\\ (0.012)\end{tabular}   & \begin{tabular}[c]{@{}l@{}}35\\ (5)\end{tabular}    & 0.96  \\ \midrule
\multirow{4}{*}{\begin{tabular}[c]{@{}l@{}}(5e4,\\  1e2,\\  5e2)\end{tabular}}  & VSBM   & -                                                       & -                                                     & -     & VDCSBM   & -                                                        & -                                                     & -     \\
                                                                                & D-VSBM & \begin{tabular}[c]{@{}l@{}}0.96 \\ (0.010)\end{tabular} & \begin{tabular}[c]{@{}l@{}}512\\ (10)\end{tabular}  & 0.96  & D-VDCSBM & \begin{tabular}[c]{@{}l@{}}0.95 \\ (0.012)\end{tabular}  & \begin{tabular}[c]{@{}l@{}}553\\ (12)\end{tabular}  & 0.97  \\
                                                                                & SSP    & -                                                       & -                                                     & -     & SCORE    & -                                                        & -                                                     & -     \\
                                                                                & D-SSP  & \begin{tabular}[c]{@{}l@{}}0.90 \\ (0.008)\end{tabular} & \begin{tabular}[c]{@{}l@{}}273\\ (11)\end{tabular}  & 0.96  & D-SCORE  & \begin{tabular}[c]{@{}l@{}}0.88\\ (0.013)\end{tabular}   & \begin{tabular}[c]{@{}l@{}}394\\ (12)\end{tabular}  & 0.97  \\ \midrule
\multirow{4}{*}{\begin{tabular}[c]{@{}l@{}}(5e5,\\  1e3,\\  5e3)\end{tabular}}  & VSBM   & -                                                       & -                                                     & -     & VDCSBM   & -                                                        & -                                                     & -     \\
                                                                                & D-VSBM & \begin{tabular}[c]{@{}l@{}}0.96 \\ (0.010)\end{tabular} & \begin{tabular}[c]{@{}l@{}}5590\\ (35)\end{tabular} & 0.94  & D-VDCSBM & \begin{tabular}[c]{@{}l@{}}0.982 \\ (0.010)\end{tabular} & \begin{tabular}[c]{@{}l@{}}6024\\ (32)\end{tabular} & 0.97  \\
                                                                                & SSP    & -                                                       & -                                                     & -     & SCORE    & -                                                        & -                                                     & -     \\
                                                                                & D-SSP  & \begin{tabular}[c]{@{}l@{}}0.90 \\ (0.009)\end{tabular} & \begin{tabular}[c]{@{}l@{}}3025\\ (20)\end{tabular} & 0.94  & D-SCORE  & \begin{tabular}[c]{@{}l@{}}0.975 \\ (0.012)\end{tabular} & \begin{tabular}[c]{@{}l@{}}4520\\ (25)\end{tabular} & 0.97  \\ \bottomrule
\end{tabular}
\caption{Result for relatively large networks. Computational time includes two parts: 1) community size selection 2) community label detection. Using LRBIC to determine the community size grows exponentially with the size of networks. Our distributed community detection method can calculate the community size as well as get a high NMI score in community detection. } 
\label{Tab:comparemethodslargen}
\end{table}

\section{\label{sec:6}Data Examples}

{In this section, we apply the proposed method to real-world networks including an airline route network and Facebook ego networks.
Since the true community labels of real application are unknown, the evaluation step is based on two aspects: first, in airline route network, we analyze the community detection results from regions' geometric and economic factors; 
second, in both applications, we analyze the AUC results by considering the recovering power of edges based on the estimated SBM model.
}

\subsection{Airline Route Network}\label{subsec:6.1}

The airline route network\footnote{https://openflights.org/data.html}  contains 67,663 routes between 3,321 airports on 548 airlines spanning the globe. The adjacency matrix $\V A$ is then constructed as follows: each airport is considered as a node; if there exists a route between two nodes $v_i$ and $v_j$, we set $A_{ij} = 1$, otherwise, $A_{ij} = 0$.  We apply the proposed method and report the results based on D-SSP.  The results from D-VSBM are similar and omitted here.

\begin{figure}[ht]
\centering
\includegraphics[width=0.9\textwidth]{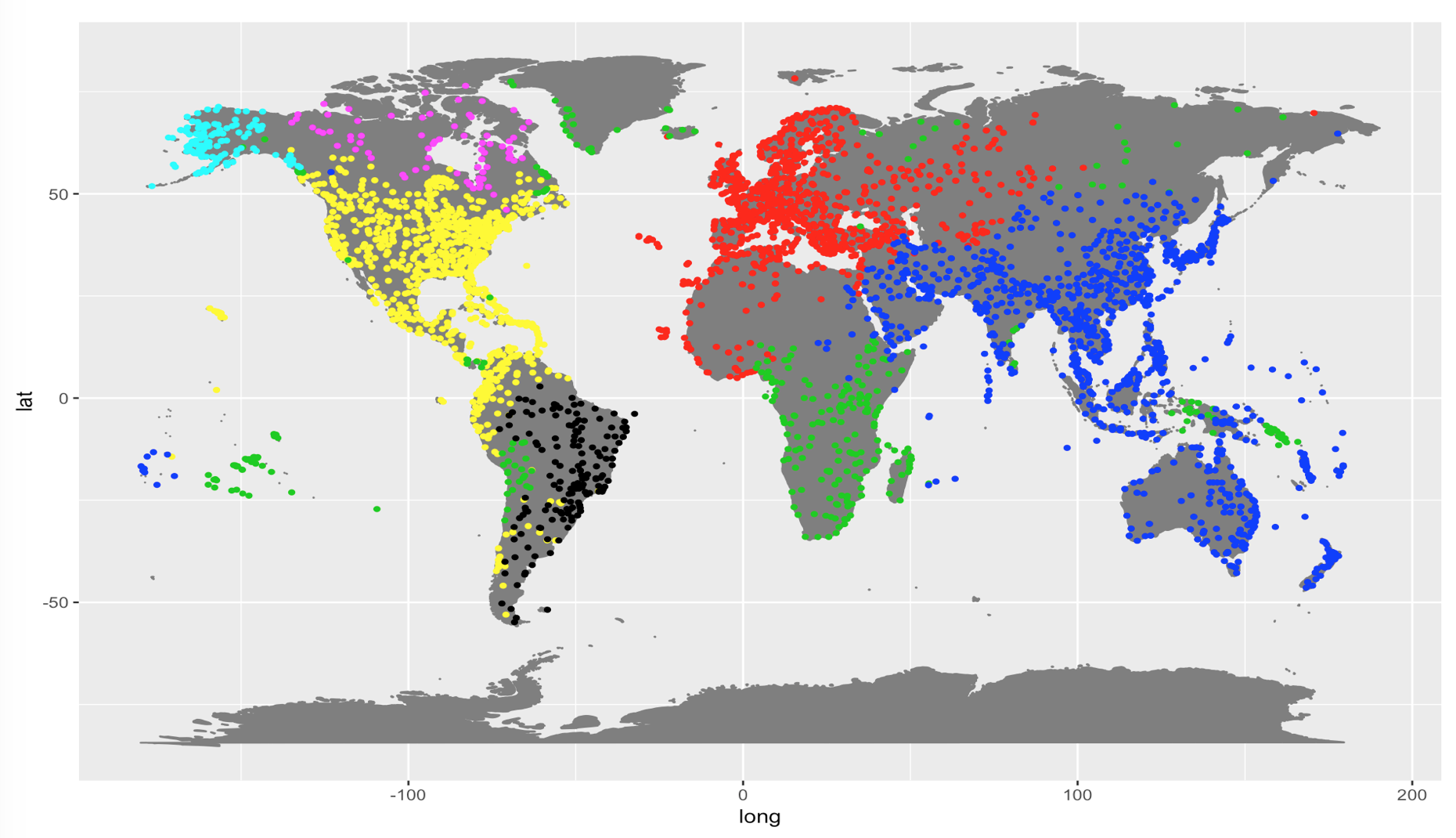}
\caption{Group detection in the airline route network; different colors indicate different estimated group labels.}
\label{fig:airport_group}
\end{figure}

\begin{figure}[!htb]
\begin{subfigure}{0.55\textwidth}
\begin{minipage}{0.5\textwidth}
\centering
\includegraphics[width=0.8\linewidth]{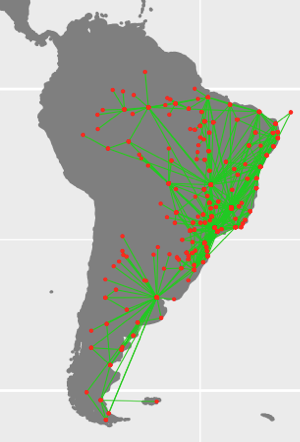}
\end{minipage}\hfill
\begin{minipage}{0.5\textwidth}
\centering
\includegraphics[width=0.88\linewidth]{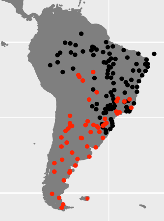}
\end{minipage}
\caption{South America (black points in Figure \ref{fig:airport_group} ) }
\end{subfigure}
\begin{subfigure}{0.45\textwidth}
\centering
\begin{minipage}{0.7\textwidth}
\centering
\includegraphics[width=1\linewidth]{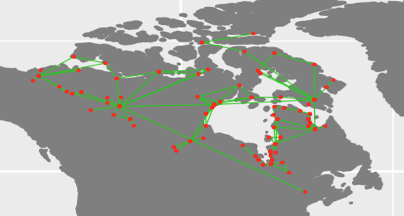}
\end{minipage}\hfill
\begin{minipage}{0.7\textwidth}
\centering
\includegraphics[width=1\linewidth]{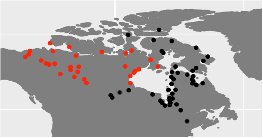}
\end{minipage}
\caption{Canada (pink points in Figure \ref{fig:airport_group})}
\end{subfigure}

\caption{In each sub-figure, one panel shows the airline routes (with red points corresponding to airports and green lines corresponding to airline routes), and the other panel shows the estimated community labels - different colors indicate different estimated community labels. {In Canada, there are three airports (in black) that are obviously mislabeled.} }\label{Fig:airport_region}
\end{figure}

\begin{figure}[!htb]
\begin{subfigure}{0.5\textwidth}
\centering
\includegraphics[width=1\linewidth]{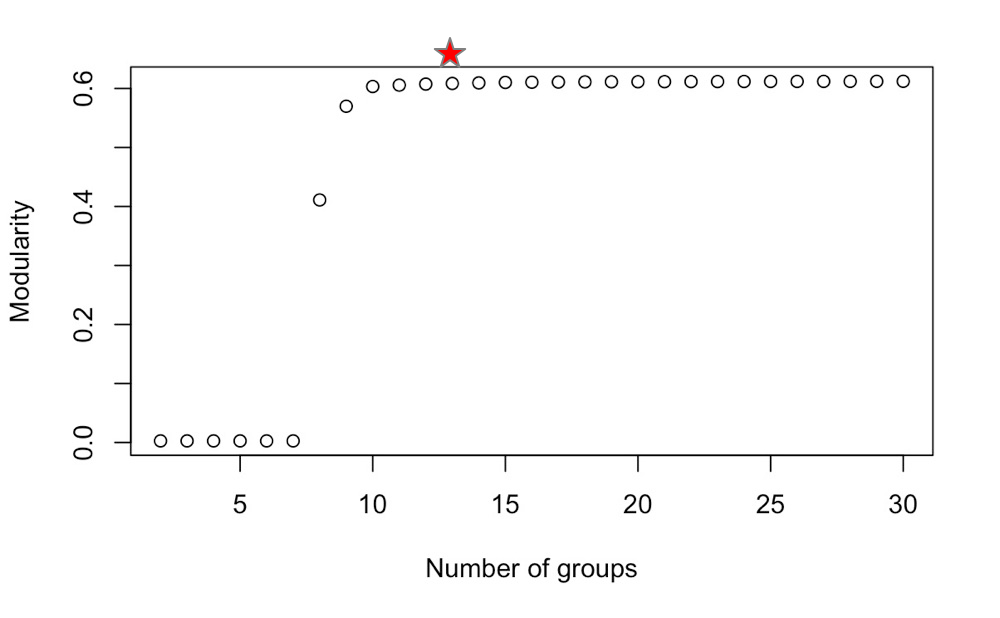}
\caption{Airline route network}
\end{subfigure}\hfill
\begin{subfigure}{0.5\textwidth}
\centering
\includegraphics[width=1\linewidth]{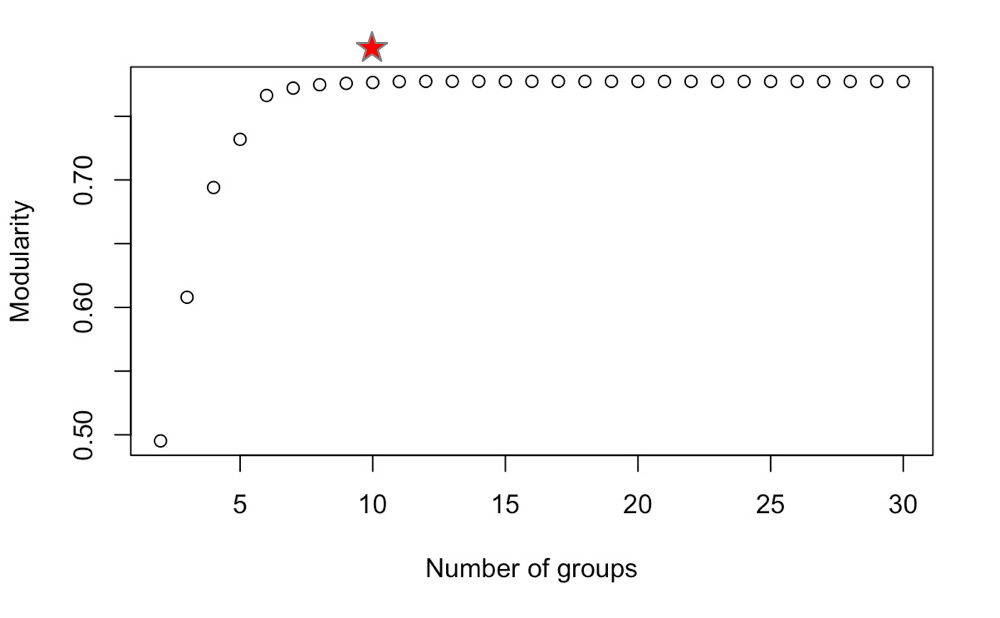}
\caption{Facebook ego network}
\end{subfigure}
\caption{Modularity under different values of $G$}
\label{fig:modularity_analysis}
\end{figure}

We first choose the number of groups $G$. Figure \ref{fig:modularity_analysis}(a) shows the modularity value under different values of $G$.  As one can see, the modularity value reaches its peak when $G=13$ and then remains almost the same as $G$ further increases.  Therefore, we divide the network into $G=13$ groups.

Figure \ref{fig:airport_group} shows the seven largest identified groups, where different colors correspond to different estimated groups. Roughly speaking, these seven groups represent seven different regions in the world. For example, the black points are mainly in South America, the pink points are mainly in Canada, while the yellow points are mainly in the United States, Mexico, and the Caribbean region. 

Community detection is then applied to each identified group. For example, the right panel of Figure \ref{Fig:airport_region}(a) shows the community detection results for the group in South America.  As one can see, the two detected communities correspond to the two centers in the left panel of Figure \ref{Fig:airport_region}(a): one in Uruguay and Argentina, and the other in Brazil, which are understandable. Another example is shown in Figure \ref{Fig:airport_region}(b). Two detected communities (lower panel) match the airline routes in the upper panel and correspond to eastern and western Canada, respectively.

Further, we evaluate the effectiveness of community detection in a predictive manner. 
Specifically, we first randomly sample a proportion of the node pairs as the testing set and mask the corresponding edge $A_{ij}$ values as zero; then we apply the proposed community detection method to obtain $\hat{\V B}_n$ which can be estimated according to the detected community labels $\hat{\V c}$ by considering the connection probability of nodes both within and between communities.
Lastly, we evaluate the performance of $\hat{\V B}_n$ on the testing set using the area under the receiver operating characteristic curve (AUC).
{The left panel in Figure \ref{fig:auc} shows how the AUC value changes when we vary the proportion of masked node pairs.  As we can see, overall, the AUC values are pretty high and stable.  As the proportion of masked node pairs increases, the AUC value decreases, but not by a lot.  We can also see that the AUC of D-SSP is only slightly lower than that of SSP, indicating again that with data splitting, the proposed distributed community detection method does not degrade the community detection performance.}


\subsection{Ego Networks} \label{subsec:6.2}

Ego networks \citep{leskovec2012learning, snapnets} are often created from user surveys in social media, such as Facebook, Twitter, or Google$+$.  Each ego person is considered as a primary node, and the ego's friends are all considered as nodes in the ego network.  If two nodes in the ego network are friends with each other, a link is added between these two nodes. 
For example, the Facebook ego network we consider here 
contains 4,039 nodes and 88,234 edges, corresponding to 10 ego people (see Figure \ref{fig:egofb}).

\begin{figure}[!htb]
\begin{subfigure}{0.5\textwidth}
\centering
\includegraphics[width=1\linewidth]{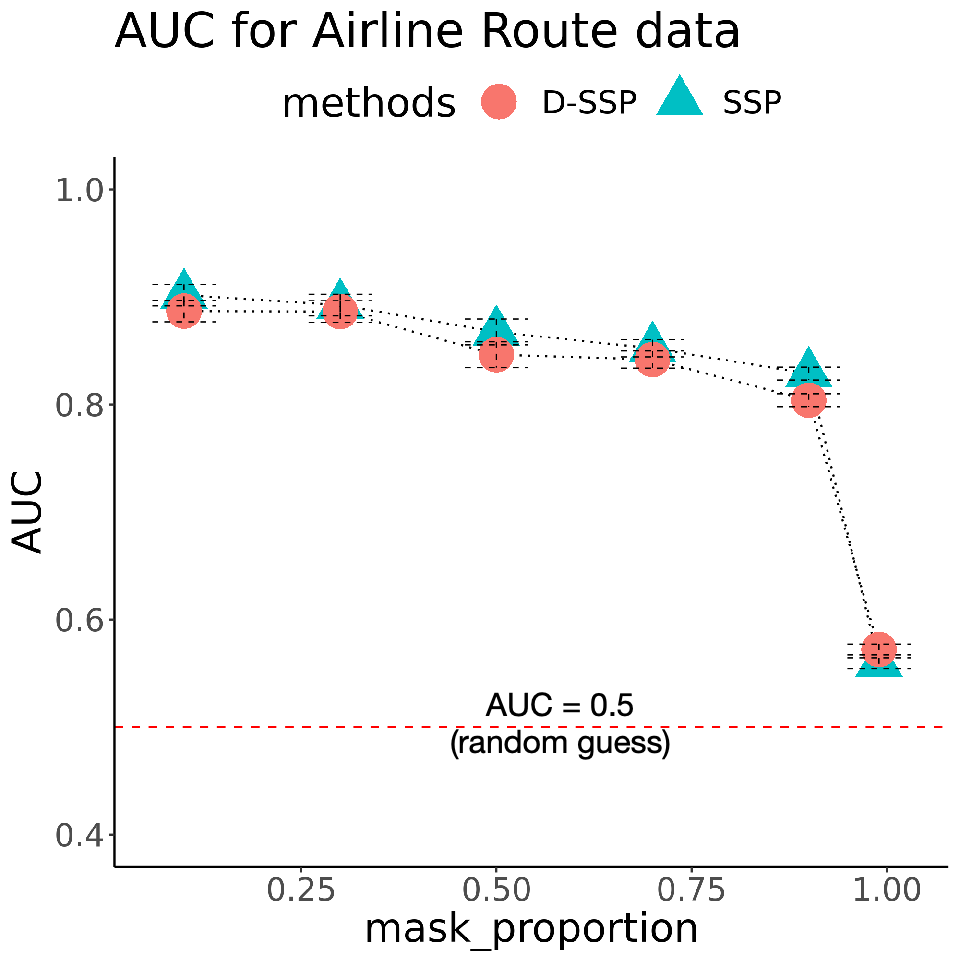}
\end{subfigure}\hfill
\begin{subfigure}{0.5\textwidth}
\centering
\includegraphics[width=1\linewidth]{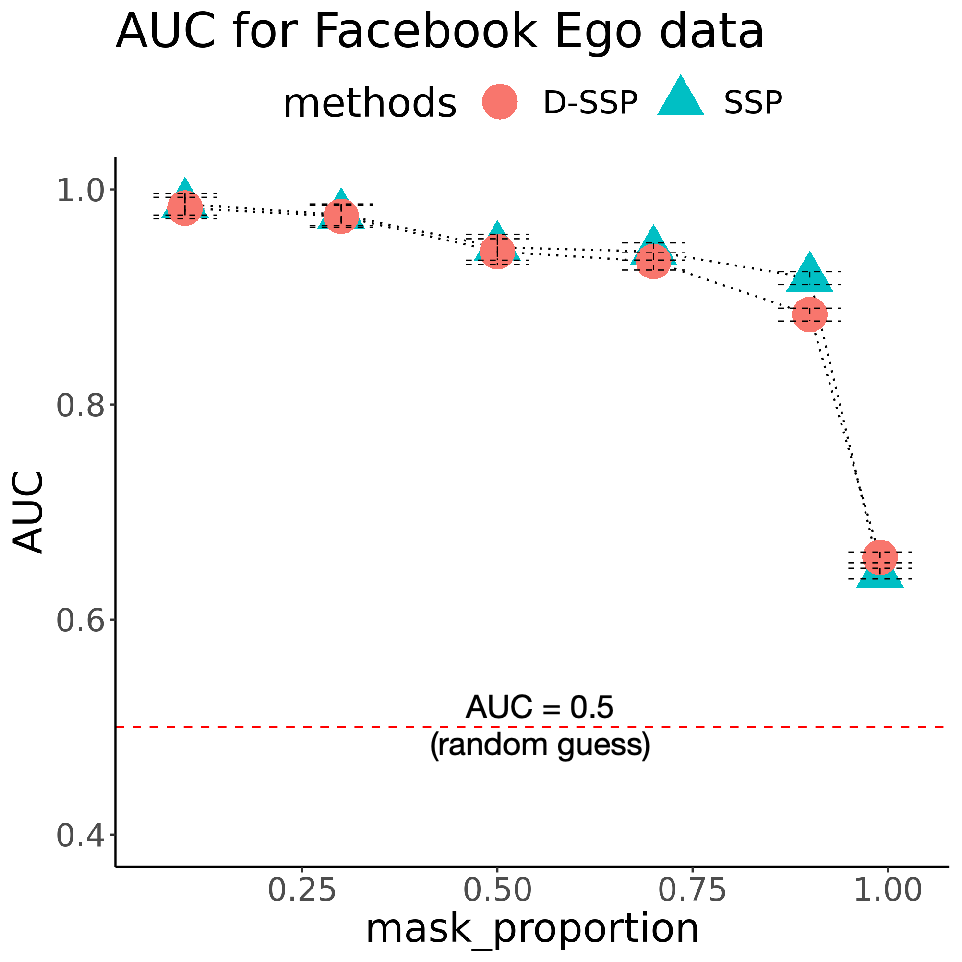}
\end{subfigure}
\caption{AUC under different proportions of masked node pairs. {The x-axis for masked proportion is set as 0.1, 0.3, 0.5, 0.7, 0.99 respectively.}}
\label{fig:auc}
\end{figure}

Since there are 10 ego people in this Facebook ego network, it is reasonable to set the number of groups $G$ as 10.  The plot of modularity for the ego network (Figure \ref{fig:modularity_analysis}(b)), which shows the modularity value under different values of $G$, confirms the conjecture. As one can see, the modularity value reaches its peak at $G = 10$ and stays at about the same value for large $G$ values.  After setting the number of groups as 10, we apply the D-SSP method to the Facebook ego network to obtain the estimated community labels and link probability matrix.  Similar to the previous subsection, the results are evaluated using the AUC on the testing set (see Figure \ref{fig:auc}).

\begin{table}[!ht]
\centering
\begin{tabular}{ | c | c | c | c | p{1.5cm} | p{1.5cm} | p{1.2cm}| p{1.2cm} |}
\hline
Network & Egos & Nodes & Edges & AUC (mask $10\%$) (D-SSP)  & Time (D-SSP) &  AUC (mask $10\%$) (SSP) & Time (SSP)  \\
[0.5ex] 
\hline
Facebook & 10 & 4,039 & 88,234 & 0.985 & 44(s) & 0.986 & 214(s) \\
\hline
Twitter & 42 & 5,816 & 83,694 & 0.991 & 34(s) & 0.989 & 420(s) \\
\hline
Google$+$ & 8 & 6,382 & 325,819 & 0.941 & 101(s) & 0.954 & 761(s) \\
\hline
Twitter & 142 & 81,306 & 1,342,296 & 0.989 & 1091(s) & ------ & ------ \\
\hline
Google$+$ & 132 & 107,614 & 13,673,453 & 0.974 & 5030(s) & ------ & ------ \\
\hline
\end{tabular}
\caption{Numerical results for ego networks; all experiments are conducted via a Dell R7425 machine with dual processor AMD Epyc 32 core 2.2 GHZ with 8GB of RAM}
\label{table:Realdata}
\end{table}

Further, we have also considered a Twitter ego network and a Google$+$ ego network.  Since these two networks are relatively large (with 142 and 132 ego people respectively), the SSP method can not be applied due to the computational cost.  We thus apply the SSP method only on a subset of the ego network corresponding to a randomly sampled subset of the ego people.  The results are shown in Table \ref{table:Realdata}.  It can be seen that the proposed distributed community detection method D-SSP has comparable performance to SSP (when SSP can be applied) in terms of the AUC, but significantly reducing the computational cost.



\section{\label{sec:7}Conclusion and Future Work}

In this paper, we have proposed a novel distributed community detection method for large networks with a grouped community structure.  The method is easy to implement using the current technology in modularity optimization and community detection.  We have also established both weak and strong consistency of the proposed method for both group and community detection. Numerical results show that the proposed distributed community detection method achieves similar detection performance in comparison to community detection methods without data splitting, but largely reduces the computational cost.

There are several directions we plan to extend our work.  The number of groups $G$ is currently assumed fixed as $n$ grows in our theoretical analysis.  Using techniques developed in \citet{choi2012stochastic} and \citet{rohe2011spectral}, we may relax this assumption.  Another interesting direction is to generalize the current work to incorporate the dynamic representation of time-varying networks with evolving community structures \citep{sarkar2006dynamic,berger2006framework}. This is challenging and requires further investigation.

\vskip 0.2in
\bibliography{sampbib}

\begin{thebibliography}{33}
\providecommand{\natexlab}[1]{#1}
\providecommand{\url}[1]{\texttt{#1}}
\expandafter\ifx\csname urlstyle\endcsname\relax
  \providecommand{\doi}[1]{doi: #1}\else
  \providecommand{\doi}{doi: \begingroup \urlstyle{rm}\Url}\fi

\bibitem[Abbe(2017)]{Abbe2017CommunityDA}
Emmanuel Abbe.
\newblock Community detection and stochastic block models: recent developments.
\newblock \emph{J. Mach. Learn. Res.}, 18:\penalty0 177:1--177:86, 2017.

\bibitem[Amini et~al.(2013)Amini, Chen, Bickel, and Levina]{amini2013pseudo}
Arash~A Amini, Aiyou Chen, Peter~J Bickel, and Elizaveta Levina.
\newblock Pseudo-likelihood methods for community detection in large sparse
  networks.
\newblock \emph{The Annals of Statistics}, 41\penalty0 (4):\penalty0
  2097--2122, 2013.

\bibitem[Berger-Wolf and Saia(2006)]{berger2006framework}
Tanya~Y Berger-Wolf and Jared Saia.
\newblock A framework for analysis of dynamic social networks.
\newblock In \emph{Proceedings of the 12th ACM SIGKDD international conference
  on Knowledge discovery and data mining}, pages 523--528. ACM, 2006.

\bibitem[Bickel and Chen(2009)]{bickel2009nonparametric}
Peter~J Bickel and Aiyou Chen.
\newblock A nonparametric view of network models and newman--girvan and other
  modularities.
\newblock \emph{Proceedings of the National Academy of Sciences}, pages
  pnas--0907096106, 2009.

\bibitem[Blondel et~al.(2008)Blondel, Guillaume, Lambiotte, and
  Lefebvre]{blondel2008fast}
Vincent~D Blondel, Jean-Loup Guillaume, Renaud Lambiotte, and Etienne Lefebvre.
\newblock Fast unfolding of communities in large networks.
\newblock \emph{Journal of statistical mechanics: theory and experiment},
  2008\penalty0 (10):\penalty0 P10008, 2008.

\bibitem[Choi et~al.(2012)Choi, Wolfe, and Airoldi]{choi2012stochastic}
David~S Choi, Patrick~J Wolfe, and Edoardo~M Airoldi.
\newblock Stochastic blockmodels with a growing number of classes.
\newblock \emph{Biometrika}, 99\penalty0 (2):\penalty0 273--284, 2012.

\bibitem[Clauset et~al.(2004)Clauset, Newman, and Moore]{clauset2004finding}
Aaron Clauset, Mark~EJ Newman, and Cristopher Moore.
\newblock Finding community structure in very large networks.
\newblock \emph{Physical review E}, 70\penalty0 (6):\penalty0 066111, 2004.

\bibitem[Daudin et~al.(2008)Daudin, Picard, and Robin]{daudin2008mixture}
J-J Daudin, Franck Picard, and St{\'e}phane Robin.
\newblock A mixture model for random graphs.
\newblock \emph{Statistics and computing}, 18\penalty0 (2):\penalty0 173--183,
  2008.

\bibitem[Egghe and Rousseau(1990)]{egghe1990introduction}
Leo Egghe and Ronald Rousseau.
\newblock \emph{Introduction to informetrics: Quantitative methods in library,
  documentation and information science}.
\newblock Elsevier Science Publishers, 1990.

\bibitem[Fortunato and Barthelemy(2007)]{fortunato2007resolution}
Santo Fortunato and Marc Barthelemy.
\newblock Resolution limit in community detection.
\newblock \emph{Proceedings of the National Academy of Sciences}, 104\penalty0
  (1):\penalty0 36--41, 2007.

\bibitem[Fortunato and Hric(2016)]{fortunato2016community}
Santo Fortunato and Darko Hric.
\newblock Community detection in networks: A user guide.
\newblock \emph{Physics Reports}, 659:\penalty0 1--44, 2016.

\bibitem[Guimera et~al.(2004)Guimera, Sales-Pardo, and
  Amaral]{guimera2004modularity}
Roger Guimera, Marta Sales-Pardo, and Lu{\'\i}s A~Nunes Amaral.
\newblock Modularity from fluctuations in random graphs and complex networks.
\newblock \emph{Physical Review E}, 70\penalty0 (2):\penalty0 025101, 2004.

\bibitem[Handcock et~al.(2007)Handcock, Raftery, and
  Tantrum]{handcock2007model}
Mark~S Handcock, Adrian~E Raftery, and Jeremy~M Tantrum.
\newblock Model-based clustering for social networks.
\newblock \emph{Journal of the Royal Statistical Society: Series A (Statistics
  in Society)}, 170\penalty0 (2):\penalty0 301--354, 2007.

\bibitem[Holland et~al.(1983)Holland, Laskey, and
  Leinhardt]{holland1983stochastic}
Paul~W Holland, Kathryn~Blackmond Laskey, and Samuel Leinhardt.
\newblock Stochastic blockmodels: First steps.
\newblock \emph{Social networks}, 5\penalty0 (2):\penalty0 109--137, 1983.

\bibitem[Jin et~al.(2015)]{jin2015fast}
Jiashun Jin et~al.
\newblock Fast community detection by score.
\newblock \emph{The Annals of Statistics}, 43\penalty0 (1):\penalty0 57--89,
  2015.

\bibitem[Karrer and Newman(2011)]{karrer2011stochastic}
Brian Karrer and Mark~EJ Newman.
\newblock Stochastic blockmodels and community structure in networks.
\newblock \emph{Physical review E}, 83\penalty0 (1):\penalty0 016107, 2011.

\bibitem[Lei et~al.(2015)Lei, Rinaldo, et~al.]{lei2015consistency}
Jing Lei, Alessandro Rinaldo, et~al.
\newblock Consistency of spectral clustering in stochastic block models.
\newblock \emph{The Annals of Statistics}, 43\penalty0 (1):\penalty0 215--237,
  2015.

\bibitem[Leskovec and Krevl(2014)]{snapnets}
Jure Leskovec and Andrej Krevl.
\newblock {SNAP Datasets}: {Stanford} large network dataset collection.
\newblock \url{http://snap.stanford.edu/data}, June 2014.

\bibitem[Leskovec and Mcauley(2012)]{leskovec2012learning}
Jure Leskovec and Julian~J Mcauley.
\newblock Learning to discover social circles in ego networks.
\newblock In \emph{Advances in neural information processing systems}, pages
  539--547, 2012.

\bibitem[Mori and Malik(2003)]{mori2003recognizing}
Greg Mori and Jitendra Malik.
\newblock Recognizing objects in adversarial clutter: Breaking a visual
  captcha.
\newblock In \emph{Computer Vision and Pattern Recognition, 2003. Proceedings.
  2003 IEEE Computer Society Conference on}, volume~1, pages I--I. IEEE, 2003.

\bibitem[Newman(2006)]{newman2006modularity}
Mark~EJ Newman.
\newblock Modularity and community structure in networks.
\newblock \emph{Proceedings of the national academy of sciences}, 103\penalty0
  (23):\penalty0 8577--8582, 2006.

\bibitem[Newman(2012)]{newman2012communities}
Mark~EJ Newman.
\newblock Communities, modules and large-scale structure in networks.
\newblock \emph{Nature physics}, 8\penalty0 (1):\penalty0 25, 2012.

\bibitem[Newman and Girvan(2004)]{newman2004finding}
Mark~EJ Newman and Michelle Girvan.
\newblock Finding and evaluating community structure in networks.
\newblock \emph{Physical review E}, 69\penalty0 (2):\penalty0 026113, 2004.

\bibitem[Qin and Rohe(2013)]{qin2013regularized}
Tai Qin and Karl Rohe.
\newblock Regularized spectral clustering under the degree-corrected stochastic
  blockmodel.
\newblock In \emph{Advances in Neural Information Processing Systems}, pages
  3120--3128, 2013.

\bibitem[Rohe et~al.(2011)Rohe, Chatterjee, and Yu]{rohe2011spectral}
Karl Rohe, Sourav Chatterjee, and Bin Yu.
\newblock Spectral clustering and the high-dimensional stochastic blockmodel.
\newblock \emph{The Annals of Statistics}, 39\penalty0 (4):\penalty0
  1878--1915, 2011.

\bibitem[Sarkar and Moore(2006)]{sarkar2006dynamic}
Purnamrita Sarkar and Andrew~W Moore.
\newblock Dynamic social network analysis using latent space models.
\newblock In \emph{Advances in Neural Information Processing Systems}, pages
  1145--1152, 2006.

\bibitem[Snijders and Nowicki(1997)]{snijders1997estimation}
Tom~AB Snijders and Krzysztof Nowicki.
\newblock Estimation and prediction for stochastic blockmodels for graphs with
  latent block structure.
\newblock \emph{Journal of classification}, 14\penalty0 (1):\penalty0 75--100,
  1997.

\bibitem[Von~Luxburg et~al.(2008)Von~Luxburg, Belkin, and
  Bousquet]{von2008consistency}
Ulrike Von~Luxburg, Mikhail Belkin, and Olivier Bousquet.
\newblock Consistency of spectral clustering.
\newblock \emph{The Annals of Statistics}, pages 555--586, 2008.

\bibitem[Wang et~al.(2020)Wang, Zhang, Liu, Zhu, and Guo]{wang2020fast}
Jiangzhou Wang, Jingfei Zhang, Binghui Liu, Ji~Zhu, and Jianhua Guo.
\newblock Fast network community detection with profile-pseudo likelihood
  methods.
\newblock \emph{arXiv preprint arXiv:2011.00647}, 2020.

\bibitem[Wang and Bickel(2017)]{wang2017likelihood}
YX~Rachel Wang and Peter~J Bickel.
\newblock Likelihood-based model selection for stochastic block models.
\newblock \emph{The Annals of Statistics}, 45\penalty0 (2):\penalty0 500--528,
  2017.

\bibitem[White and Smyth(2005)]{white2005spectral}
Scott White and Padhraic Smyth.
\newblock A spectral clustering approach to finding communities in graphs.
\newblock In \emph{Proceedings of the 2005 SIAM international conference on
  data mining}, pages 274--285. SIAM, 2005.

\bibitem[Zhao et~al.(2011)Zhao, Levina, and Zhu]{zhao2011community}
Yunpeng Zhao, Elizaveta Levina, and Ji~Zhu.
\newblock Community extraction for social networks.
\newblock \emph{Proceedings of the National Academy of Sciences}, 2011.

\bibitem[Zhao et~al.(2012)Zhao, Levina, and Zhu]{zhao2012consistency}
Yunpeng Zhao, Elizaveta Levina, and Ji~Zhu.
\newblock Consistency of community detection in networks under degree-corrected
  stochastic block models.
\newblock \emph{The Annals of Statistics}, 40\penalty0 (4):\penalty0
  2266--2292, 2012.

\end{thebibliography}

\newpage
\appendix

\section*{Appendix}
\label{app:theorem}



We first formally formulate the group detection problem for networks with the grouped community structure.  Given the total number of groups $G$, we use $\V e$ to denote a
group assignment $\V e = \{e_1,e_2,\ldots,e_n\}$, where $e_i$ is the group label for node $v_i$ and it takes value in $\{1,2,\ldots,G\}$.  
Define $\hat{\V g} = \mbox{argmax}_{\V e}{Q_{ER}(\V e)}$ as the estimated group labels. Our first goal is to show that $\hat {\V g}$ can consistently estimate the true group labels $\V g$. 
Following \cite{bickel2009nonparametric} and \cite{zhao2012consistency}, for any given group assignment $\V e$, we construct matrix $\V O(\V e) \in \mathbb{R}^{G \times G}$ as 
$$ O_{ts}(\V e) = \sum_{ij}A_{ij}I\{e_i = t,e_j = s\}, $$ 
where $I$ is the indicator function. The total degree in group $t$ under the group assignment $\V e$ can be defined as 
$$ O_t(\V e) = \sum_s O_{ts}(\V e), \qquad t=1,\dots,G. $$ 
Let $\V f(\V e) = (\frac{n_1(e)}{n}, \frac{n_2(e)}{n},...,\frac{n_G(e)}{n})^T$ denote the frequencies in each group under the group assignment $\V e$, where $n_t(\V e) = \sum_iI\{e_i = t\}$ is the number of nodes in group $t$. Also, denote the sum of degrees as $L = \sum_{ij}A_{ij}$.

We start with the general format of modularity-based criteria, which can be written as $$Q(\V e) = F(\frac{\V O(\V e)}{\mu_n}, \V f(\V e)).$$  We aim to find an \V e that maximizes the modularity.  

The key in group detection problem is to construct a ``population version" of $Q(\V e)$ such that the group labels $\V g$ can maximize the ``population version" of $Q(\V e)$. We first need to construct the ``population version" of $\V f(\V e)$ and $O(\V e)$.

We define a matrix $\V R(\V e) \in \mathbb{R}^{G\times K \times M}$ as $R_{tau}(\V e) = \frac{1}{n}\sum_iI(e_i = t,c_i = a, \theta_i = x_u)$,
which is a normalized one-hot representation of the relationship between nodes and structure including group label, community labels and degree level.
For any generic array $\V S = [S_{tau}] \in \mathbb{R}^{G \times K \times M}$, we can define a G-dimensional vector $\V h(\V S) = [h_t(\V S)]$ by
\begin{equation}
    h_t(\V S) = \sum_{au} S_{tau},
    \label{eq:h}
\end{equation}
and $G \times G$ matrix $\V H(\V S) = [H_{ts}(\V S)]$ by 
\begin{equation}
    H_{ts}(\V S) = \sum_{abuv}x_ux_vB_{ab}S_{ta}S_{sb}.
    \label{eq:H}
\end{equation}

Then we have Lemma \ref{lemma:1} as follows:

\begin{lemma}\label{lemma:1}
    $$E(f_t(\V e) | \V c, \V \theta) = h_t(\V R(\V e)),$$
$$\frac{1}{u_n}E(O_{ts}(\V e)|\V c,  \V\theta) = H_{ts}(\V R(\V e)). $$
\end{lemma}

\noindent\textbf{Proof of Lemma \ref{lemma:1}:} 
For the first equation, left part can be written as $E(f_t(\V e) | \V c, \V \theta) = f_t(\V e) = \frac{1}{n}\sum_iI(e_i = t)$ due to the definition of $f_t(\V e)$. From the construction of $\V h$ and $\V R(\V e)$ above, we have $h_t(\V R(\V e)) = \sum_{au}R_{tau}(\V e) = \sum_{au}\frac{1}{n}\sum_iI(e_i = t,c_i = a,\theta_i = x_u)$. After changing the order of two summations, we have $$\sum_{au}\frac{1}{n}\sum_iI(e_i = t,c_i = a,\theta_i = x_u) = \sum_i\frac{1}{n}\sum_{au}I(e_i = t,c_i = a, \theta_i = x_u) = \frac{1}{n}\sum_iI(e_i = t).$$

The last equality holds because community labels $\V c$ and degrees $\V\theta$ take exact one value for each $i$. Therefore, $E(f_t(\V e) |\V c, \V\theta) = h_t(\V R(\V e))$.

For the second equation, we have by definition of $O_{ts}$:

\begin{equation*}
  \begin{split}
  \frac{1}{u_n}E(O_{ts}(\V e)|\V c, \V\theta) &= \frac{1}{u_n}E(\sum_{ij}A_{ij}{I\{e_i = t, e_j = s\}}| \V c,\V\theta) \\
 & = \frac{1}{u_n}E(\sum_{ij}\sum_{abuv}A_{ij}I\{e_i = t,g_i = a, \theta_i = x_u\}I\{e_j = s,g_j = b, \theta_j = x_v\}|\V c,\V\theta).
  \end{split}
\end{equation*}

Given community labels and degrees, the adjacency matrix $\V A$ is constructed from the probability matrix $\V B_n = \rho_n \V B$. Therefore, we have:
\begin{equation*}
  \begin{split}
  &\frac{1}{u_n}E(\sum_{ij}\sum_{abuv}A_{ij}I\{e_i = t,g_i = a, \theta_i = x_u\}I\{e_j = s,g_j = b, \theta_j = x_v\}|\V c, \V\theta) \\
  &= \frac{\rho_n}{u_n}\sum_{ij}\theta_i\theta_jB_{c_i,c_j}\sum_{abuv}I\{e_i = t,g_i = a, \theta_i = x_u\}I\{e_j = s,g_j = b, \theta_j = x_v\} \\
  & = \frac{1}{n^2}\sum_{abuv}x_ux_vB_{ab}\sum_{ij}I\{e_i = t,g_i = a, \theta_i = x_u\}I\{e_j = s,g_j = b, \theta_j = x_v\} .
  \end{split}
\end{equation*}

From the construction of $\V H$ and $\V R(\V e)$, the following equations holds:
\begin{equation*}
    \begin{split}
    H_{ts}(\V R(\V e)) & =  \frac{1}{n^2}\sum_{abuv}x_ux_vB_{ab}R_{ta}(\V e)R_{sb}(\V e)\\ &= \sum_{ab}x_ux_vB_{ab} \sum_iI(e_i = t,c_i = a)\sum_jI(e_j = s,c_j = b).
    \end{split}
\end{equation*}

Therefore, $\frac{1}{u_n}E(O_{ts}(\V e)|\V c) = H_{ts}(\V R(\V e)). $  \hfill\BlackBox

From the Lemma \ref{lemma:1}, the ``population version" of $Q(\V e)$ can be defined as the expectation conditional on $\V c$ but replacing $\V R(\V e)$ by $\V T(\V e) \in \mathbb{R}^{G \times K \times M}$, where 
\begin{equation*}
    T_{tau}(\V e) = \frac{\sum_i I(e_i = t,c_i = a, \theta_i = x_u)}{\sum_iI(c_i = a, \theta_i = x_u)}\tau_{au},
\end{equation*}
it can be shown that $T_{tau}(\V e) = R_{tau}(\V e) \tau_{au} / \hat\tau_{au}$, where $\tau_{au} = \mathbb{P}(c_i = a, \theta_i = x_u)$ and $\hat{\tau}_{au} =\frac{1}{n}\sum_{i} I(c_i = a , \theta_i = x_u)$.
Therefore, the ``population version" of $Q(\V e)$ can be further denoted as $F(\V H(\V T(\V e)),\V h(\V T(\V e)))$, where we omit the constant multiplier $L$. Theorem \ref{theo:1} would indicate that the group labels \V g are the maximizer of $F(\V H(\V T(\V e)),\V h(\V T(\V e)))$.

Lemma \ref{lemma:2} is based on Bernstein's inequality which is utilized in Lemma \ref{lemma:3}. Lemma \ref{lemma:3} provides the conditions for the consistency in maximizing modularity $Q(\V e)$. 

\begin{lemma}\label{lemma:2}
    Define $\V X(\V e) \in \mathbb{R}^{G \times G}$ by $X_{ts}(\V e) = \frac{\V O(\V e)}{\mu_n} - \V H(\V R(\V e))$. Let $||\V X|| = max_{s,t}|X_{st}|$ and $|\V e - \V g| = \sum_iI(e_i \ne g_i)$. Then
\begin{equation}
    P(max_{\V e}||\V X(\V e)||_\infty \geq \varepsilon) \leq 2G^{n + 2}exp(-\frac{1}{8C}\varepsilon^2\mu_n)
    \label{eq:A1}
\end{equation}
for $\varepsilon < 3C$, where $C = max_{ab}B_{ab}$.
\begin{equation}
    P(max_{|\V e - \V g| \leq m }||\V X(\V e) - \V X(\V g)||_\infty \geq \varepsilon) \leq 2\binom{n}{m}G^{m + 2}exp(-\frac{3}{8}\varepsilon\mu_n)
    \label{eq:A2}
\end{equation}
for $\varepsilon \geq 6Cm/n$.
\begin{equation}
    P(max_{|\V e - \V g| \leq m }||\V X(\V e) - \V X(\V g)||_\infty \geq \varepsilon) \leq 2\binom{n}{m}G^{m + 2}exp(-\frac{n}{16C}\varepsilon^2\mu_n)
    \label{eq:A3}
\end{equation}
for $\varepsilon \leq 6Cm/n$.
\end{lemma}

\noindent\textbf{Proof:} See the proof of Lemma A.1. in the supplementary material of \cite{zhao2012consistency}.\hfill\BlackBox

\begin{lemma}\label{lemma:3}
    For any $Q(\V e)$ of the form $$Q(\V e) = F(\frac{\V O(\V e)}{\mu_n},\V f(\V e))$$ if $\V \pi,\V B,F$ satisfy conditions (*), (a),(b),(c), then Q is strongly consistent under stochastic block
models if $\lambda_n/(log(n)) \rightarrow \infty$ and weakly consistent when $\lambda_n \rightarrow \infty.$ 

The conditions are listed as follows:

(*) $F(\V H(\V S),\V h(\V S))$ is uniquely maximized over $\mathscr{P} = \{\V S:\V S \ge 0, \sum_{t = 1}^G S_{tau} = \tau_{au}\}$ by $\V S = \V S^*$, where $\V S^*_{tau} = \tau_{au}$ for one $t$, otherwise, $\V S^*_{t'au} = 0$ for $t' \ne t$.
$\V S^*_{\cdot\cdot u}$ has only one nonzero element in each column, and communities within group stay in the same row.

(a) $F$ is Lipschitz in its arguments;

(b) Let $\V W = \V H(\V D)$. The directional derivatives $\frac{\partial^2F}{\partial\varepsilon^2}(\V M_0 + \varepsilon(\V M_1 - \V M_0), \V t_0 + \varepsilon(\V t_1 - \V t_0))|_{\varepsilon = 0 +}$ are continuous in $(\V M_1,\V t_1)$ for all $(\V M_0,\V t_0)$ in a neighborhood of $(\V W,\V\pi)$;

(c) Let $G(\V S) = F(\V H(\V S), \V h(\V S))$.  Then on $\mathscr{P}$, $\frac{\partial G((1 - \varepsilon)\V D + \varepsilon \V S )}{\partial\varepsilon}|_{\varepsilon = 0+} < -C < 0$ for all $\V\pi, \V B$.
\end{lemma}

\noindent\textbf{Proof:} 

The proof has three steps. 

Step 1: show that the modularity $Q(\V e) = F(\frac{\V O(\V e)}{\mu_n},\V f(\V e))$ is uniformly close to the population version $F(\V H(\V T(\V e)), \V h(\V T(\V e)))$. In another word, we need to show that for $\lambda_n \rightarrow \infty$, there exist $\varepsilon_n \rightarrow 0$ such that 

\begin{equation}
    P(max_{\V e}|F(\frac{\V O(\V e)}{\mu_n},\V f(\V e)) - F(\V H(\V T(\V e)), \V h(\V T(\V e)))| < \varepsilon_n) \rightarrow 1 .
    \label{eq:step1-1}
\end{equation}

Since
\begin{equation*}
    \begin{split}
        & |F(\frac{\V O(\V e)}{\mu_n},\V f(\V e)) - F(\V H(\V T(\V e)), \V h(\V T(\V e)))| \\ & \leq  |F(\frac{\V O(\V e)}{\mu_n},\V f(\V e)) - F(\V H(\V R(\V e)), \V h(\V R(\V e)))| 
         +  |F(\V H(\V R(\V e)),\V h(\V R(\V e))) - F(\V H(\V T(\V e)), \V h(\V T(\V e)))|, 
    \end{split}
\end{equation*}

For the first item on the left hand above, since $\V h(\V R(\V e)) = f(\V e)$ and by Lipschitz continuity, 
\begin{equation}
    |F(\frac{\V O(\V e)}{\mu_n},\V f(\V e)) - F(\V H(\V R(\V e)), \V f(\V e))| \leq M_1 ||\V X (\V e)||_\infty. 
    \label{eq:step1-2}
\end{equation}
From Equation \ref{eq:A1} in Lemma \ref{lemma:2}, Equation \ref{eq:step1-2} tends to 0 uniformly for $\lambda_n \rightarrow \infty$. For the second item, 
\begin{equation}
    \begin{split}
        &|F(\V H(\V R(\V e)),\V f(\V e)) - F(\V H(\V T(\V e)), \V h(\V T(\V e)))| \\ 
        &\leq M_1||\V H(\V R(\V e)) - \V h(\V T(\V e))||_\infty + M_2||\V H(\V R(\V e)) - \V h(\V T(\V e))||_2,
    \end{split}
    \label{eq:step1-3}
\end{equation}
where $||.||_2$ is Euclidean norm for vectors. Since for $\lambda_n \rightarrow \infty$, $R_{ta}(\V e) \rightarrow T_{ta}(\V e) \ \ a.s.$, Equation \ref{eq:step1-3}  converges to 0 uniformly too. Thus, Equation \ref{eq:step1-1} holds.

Step 2: show that the weak consistency of group assignments holds. Since $\V T(\V g) \in \mathscr{P}$ and has only one nonzero element in each column. By continuity and condition $(*)$, there exists $\delta_n \rightarrow 0$, if $|\V e - \V g|/n = \sum_iI(e_i \ne g_i)/n \geq \delta_n$, then 
\begin{equation*}
    F(\V H(\V T(\V g)), \V h(\V T(\V g))) - F(\V H(\V T(\V e)), \V h(\V T(\V e))) > 2\varepsilon_n.
\end{equation*} 

From Equation \ref{eq:step1-1} in step 1, it is can be shown that
\begin{equation}
    \begin{split}
        &P(max_{\{\V e: |\V e - \V g|/n\geq \delta_n \}}F(\frac{\V O(\V e)}{\mu_n},\V f(\V e)) < F(\frac{\V O(\V g)}{\mu_n},\V f(\V g))) \\
        & \geq P(|max_{\{ \V e: |\V e - \V g|/n\geq \delta_n\} }F(\frac{\V O(\V e)}{\mu_n},\V f(\V e)) - max_{\{\V e: |\V e - \V g|/n\geq \delta_n \} }F(\V H(\V T(\V e)),\V h(\V T(\V e)))|<\varepsilon_n, \\
        & |F(\frac{\V O(\V g)}{\mu_n},\V f(\V g)) - F(\V H(\V T(\V e)),\V h(\V T(\V e)))| <\varepsilon_n ) \rightarrow 1
    \end{split}
    \label{eq:step2-1}
\end{equation}

The Equation \ref{eq:step2-1} implies that $P(|\V e - \V g|/n < \delta_n) \rightarrow 1$, and weak consistency of group assignment holds.

Step 3: show that the strong consistency of group assignments holds. See the step 3 of the proof of of Theorem 4.1 in the \cite{zhao2012consistency}. \hfill\BlackBox

Now we provide the proof of Theorem \ref{theo:1}.

\noindent\textbf{Proof of Theorem \ref{theo:1}:}

Under the block assumption $\theta_i \equiv 1$, 
the population version of $Q_{ER}(\V e)$ is $F(\V H(\V S),\V h(\V S)) = \sum_t(H_{tt}(\V S) - B_0 h_t^2(\V S))$ up to a constant multiplier, where $\V S \in \mathbb{R}^{G \times K}$. Therefore, it satisfies conditions (a),(b),(c) and we only need to show condition (*) is satisfied as well. 

For $\V S \in \mathscr{P}$, we have the following equation:
\begin{equation}
    \begin{split}
        &\sum^G_t(H_{tt}(\V S) - B_0h^2_t(\V S)) + \sum^G_{t\ne s}(H_{ts}(\V S) - B_0h_t(\V S)h_s(\V S)) \\
        & = \sum^G_{t,s}H_{ts}(\V S) - B_0(\sum^G_{t}h_t(\V S))^2 \\
        & = 1 - 1^2 = 0.
    \end{split}
    \label{eq:trick1}
\end{equation}
The second last equality holds due to Lemma \ref{lemma:1} and the fact that $\sum_t f_t = 1$ and $\frac{\sum_{ts}O_{ts}}{L} = 1$. Also, we construct $\V \Delta \in \mathbb{R}^{G \times G}$ by:
\begin{equation}
    \Delta_{ts} = (I(t = s) - 1/2)\times 2.
    \label{eq:trick2}
\end{equation}

From Equation \ref{eq:trick1} and \ref{eq:trick2}, the population version $F(\V H(\V S),\V h(\V S))$ can be rewritten as:
\begin{equation*}
    \begin{split}
        F(\V H(\V S),\V h(\V S)) &= \sum^G_t(H_{tt}(\V S) - B_0h_t^2(\V S)) \\
        &= \frac{1}{2}\sum^G_{t,s}\Delta_{ts}(H_{ts}(\V S) - B_0h_t(\V S)h_s(\V S)).
    \end{split}
\end{equation*}

From the Definition of $h$ and $H$ in equation \ref{eq:h} and \ref{eq:H}, we get: 
\begin{equation*}
    \begin{split}
        &\frac{1}{2}\sum^G_{t,s}\Delta_{ts}(H_{ts}(\V S) - B_0h_t(\V S)h_s(\V S)) \\
        =& \frac{1}{2}\sum^G_{t,s}\Delta_{ts}(\sum^K_{a,b}B_{ab}S_{ta}S_{sb} - B_0\sum^K_aS_{ta}\sum^K_bS_{sb})\\
        =& \frac{1}{2}\sum^G_{t,s}\sum^K_{a,b}S_{ta}S_{sb}\Delta_{ts}(B_{ab} - B_0).
    \end{split}
\end{equation*}

Construct $\V \Delta^{\V c} \in \mathbb{R}^{K \times K}$ by: 
\begin{equation*}
     \Delta^{\V c}_{ab}= \begin{cases}
    1, community \ a,\ b \ belong \ to \ same \ group \\
    -1,  O.W.
    \end{cases}
\end{equation*}

From the conditions in Condition \ref{cond:1}, the following inequality always holds for any $t,s,a,b$:
$$\Delta_{ts}(B_{ab} - B_0) \le \Delta^{\V c}_{ab}(B_{ab} - B_0).$$ Therefore, population version of modularity satisfies:
\begin{equation*}
    \begin{split}
        F(\V H(\V S),\V h(\V S)) &= \frac{1}{2}\sum^G_{t,s}\sum^K_{a,b}S_{ta}S_{sb}\Delta_{ts}(B_{ab} - B_0) \\
        &\le \frac{1}{2}\sum^G_{t,s}\sum^K_{a,b}S_{ta}S_{sb}\Delta^{\V c}_{ab}(B_{ab} - B_0).
    \end{split}
\end{equation*}
We can change the order of summations in last item above:
\begin{equation*}
    \begin{split}
        &\frac{1}{2}\sum^G_{t,s}\sum^K_{a,b}S_{ta}S_{sb}\Delta^{\V c}_{ab}(B_{ab} - B_0) = \frac{1}{2}\sum^K_{a,b}\Delta^{\V c}_{ab}(B_{ab} - B_0)\sum^G_{t}S_{ta}\sum^G_{s}S_{sb} \\
        & = \frac{1}{2}\sum^K_{a,b}\pi_a\pi_b\Delta^{\V c}_{ab}(B_{ab} - B_0) =  F(\V H(\V S^*),\V h(\V S^*)).
    \end{split}
\end{equation*}

The second last equality holds because $\V S \in \mathscr{P}$. Therefore, we have $F(\V H(\V S),\V h(\V S)) \le F(\V H(\V S^*),\V h(\V S^*))$ (up to row permutation of $\V S^*$). Therefore, condition $(*)$ is also satisfied. From the result in Lemma \ref{lemma:3}, Theorem \ref{theo:1} holds. \hfill\BlackBox

\noindent\textbf{Proof of Theorem \ref{theo:1dc}:}

For the degree-corrected model, we follow a similar strategy in Proof of Theorem \ref{theo:1}. The population version of $Q_{NGM}$ can be denoted as 
$$
F(H(\V S)) = \sum_{t} (\frac{H_{tt}}{\tilde{B_0}} - (\frac{H_t}{\tilde{B_0}})^2 )
$$
we need to show condition (*) is satisfied. From equation, 
$$
\sum_{t} (\frac{H_{tt}}{\tilde{B_0}} - (\frac{H_t}{\tilde{B_0}})^2) + \sum_{t \ne s} (\frac{H_{ts}}{\tilde{B_0}} - \frac{H_tH_s}{\tilde{B^2_0}}) = \sum_{ts}\frac{H_{ts}}{\tilde{B_0}} - (\sum_t \frac{H_t}{\tilde{B_0}})^2 = 0,
$$
we can show 
\begin{equation}
    \begin{split}
        F(H(\V S)) = \frac{1}{2} \Delta_{ts}(\frac{H_{ts}}{\tilde{B_0}} - (\sum_t \frac{H_t}{\tilde{B_0}})^2)
    \end{split}
\end{equation}

From the definition of $\V H$ in Equation \ref{eq:H}, it can be shown that 
$$
H_{ts} = \sum_{ab}\tilde{S}_{ta}\tilde{S}_{sb}B_{ab},
$$
and 
$$
H_t = \sum_{ac}\tilde{S}_{ta}\tilde{\pi}_cB_{ac}
$$
where $\tilde{S}_{ta} = \sum_u x_u S_{tau}$.
Hence, we obtain 

\begin{equation}
    \begin{split}
        F(H(\V S)) &= \frac{1}{2}\sum_{ts} \Delta_{ts}(\frac{\sum_{ab}\tilde{S}_{ta}\tilde{S}_{sb}B_{ab}}{\tilde{B_0}} -  \frac{\sum_{ac}\tilde{S}_{ta}\tilde{\pi}_cB_{ac} \sum_{bd}\tilde{S}_{sb}\tilde{\pi}_dB_{bd}}{\tilde{B_0}^2}) \\
        & = \frac{1}{2}\sum_{ts}\sum_{ab} \tilde{S}_{ta}\tilde{S}_{sb} \Delta_{ts}(\frac{B_{ab}}{\tilde{B_0}} -  \frac{(\sum_{c}\tilde{\pi}_cB_{ac}) (\sum_{d}\tilde{\pi}_dB_{bd})}{\tilde{B_0}^2}).
    \end{split}
\end{equation}
From the condition \ref{cond:1dc}, it satisfies that 
\begin{equation}
    \begin{split}
        &\frac{1}{2}\sum_{ts}\sum_{ab} \tilde{S}_{ta}\tilde{S}_{sb} \Delta_{ts}(\frac{B_{ab}}{\tilde{B_0}} -  \frac{(\sum_{c}\tilde{\pi}_cB_{ac}) (\sum_{d}\tilde{\pi}_dB_{bd})}{\tilde{B_0}^2}) \\
        \le & \frac{1}{2}\sum_{ts}\sum_{ab} \tilde{S}_{ta}\tilde{S}_{sb} \Delta_{ab}^{c}(\frac{B_{ab}}{\tilde{B_0}} -  \frac{(\sum_{c}\tilde{\pi}_cB_{ac}) (\sum_{d}\tilde{\pi}_dB_{bd})}{\tilde{B_0}^2}) \\
        =& F(H(\V S^*))
    \end{split}
\end{equation}
Therefore, condition $(*)$ is satisfied (up to permutation). From the result in Lemma \ref{lemma:3}, Theorem \ref{theo:1dc} holds. \hfill\BlackBox

\noindent\textbf{Proof of Theorem \ref{theo:2}:}

For weak consistency, given any $\varepsilon >  0$:
\begin{equation*}
    P(\frac{1}{n}\sum_{i =1}^n 1(\hat{c}_i \ne c_i) \ge \varepsilon) = P(\{\frac{1}{n}\sum_{i =1}^n 1(\hat{c}_i \ne c_i) \ge \varepsilon\}, \hat{\V g} = \V g)  +  P(\{\frac{1}{n}\sum_{i =1}^n 1(\hat{c}_i \ne c_i) \ge \varepsilon\}, \hat{\V g} \ne \V g)
\end{equation*}

The second item $P(\{\frac{1}{n}\sum_{i =1}^n 1(\hat{c}_i \ne c_i) \ge \varepsilon\}, \hat{\V g} \ne \V g)$ tends to zero because of the strong consistency of group labels $\hat{\V g}$.

The first item can be bounded by considering the event under each group. I.e, for nodes in group $t$, $\{\frac{1}{n}\sum_{k = 1}^{n_t}1(\hat{c}^{sub}_{t,k} \ne c^{sub}_{t,k}) \ge \varepsilon , \hat{\V g} =\V g \}$ can lead to $\{\frac{1}{n}\sum_{i =1}^n 1(\hat{c}_i \ne c_i) \ge \varepsilon, \hat{\V g} = \V g \}$. Therefore, we have: 
\begin{equation*}
    \begin{split}
       P(\{\frac{1}{n}\sum_{i =1}^n 1(\hat{c}_i \ne c_i) \ge \varepsilon\}, \hat{\V g} = \V g)  &\le P(\cup_{t = 1,2,..,G} \{\frac{1}{n}\sum_{k = 1}^{n_t}1(\hat{c}^{sub}_{t,k} \ne c^{sub}_{t,k}) \ge \varepsilon, \hat{\V g} =\V g \}) \\
       & =  P(\cup_{t = 1,2,..,G} \{\frac{1}{n_t}\sum_{k = 1}^{n_t}1(\hat{c}^{sub}_{t,k} \ne c^{sub}_{t,k}) \ge \frac{\varepsilon}{\pi_t}, \hat{\V g} = \V g \}) \\
       &\le \sum_{t = 1}^G  P( \{\frac{1}{n_t}\sum_{k = 1}^{n_t}1(\hat{c}^{sub}_{t,k} \ne c^{sub}_{t,k}) \ge \frac{\varepsilon}{\pi_t}, \hat{\V g} = \V g \}).
    \end{split}
\end{equation*}
The last part tends to zero for each $t$ due to the weak consistency of the community labels $\hat{\V c}^{sub}_t, t = 1,2,...,G$.

For strong consistency:
\begin{equation*}
    \begin{split}
        P(\hat{\V c} \ne \V c) &= P(\hat{\V g} = \V g, \hat{\V c} \ne \V c) + P(\hat{\V g} \ne \V g, \hat{\V c} \ne \V c)\\ & \rightarrow 0 + 0 = 0.
    \end{split}
\end{equation*}
The first item goes to zero from the strong consistency of the estimated community labels in each correct group; the second item goes to zero from the strong consistency of the estimated group label. \hfill\BlackBox

\begin{figure}[!htb]
\includegraphics [width=10cm] {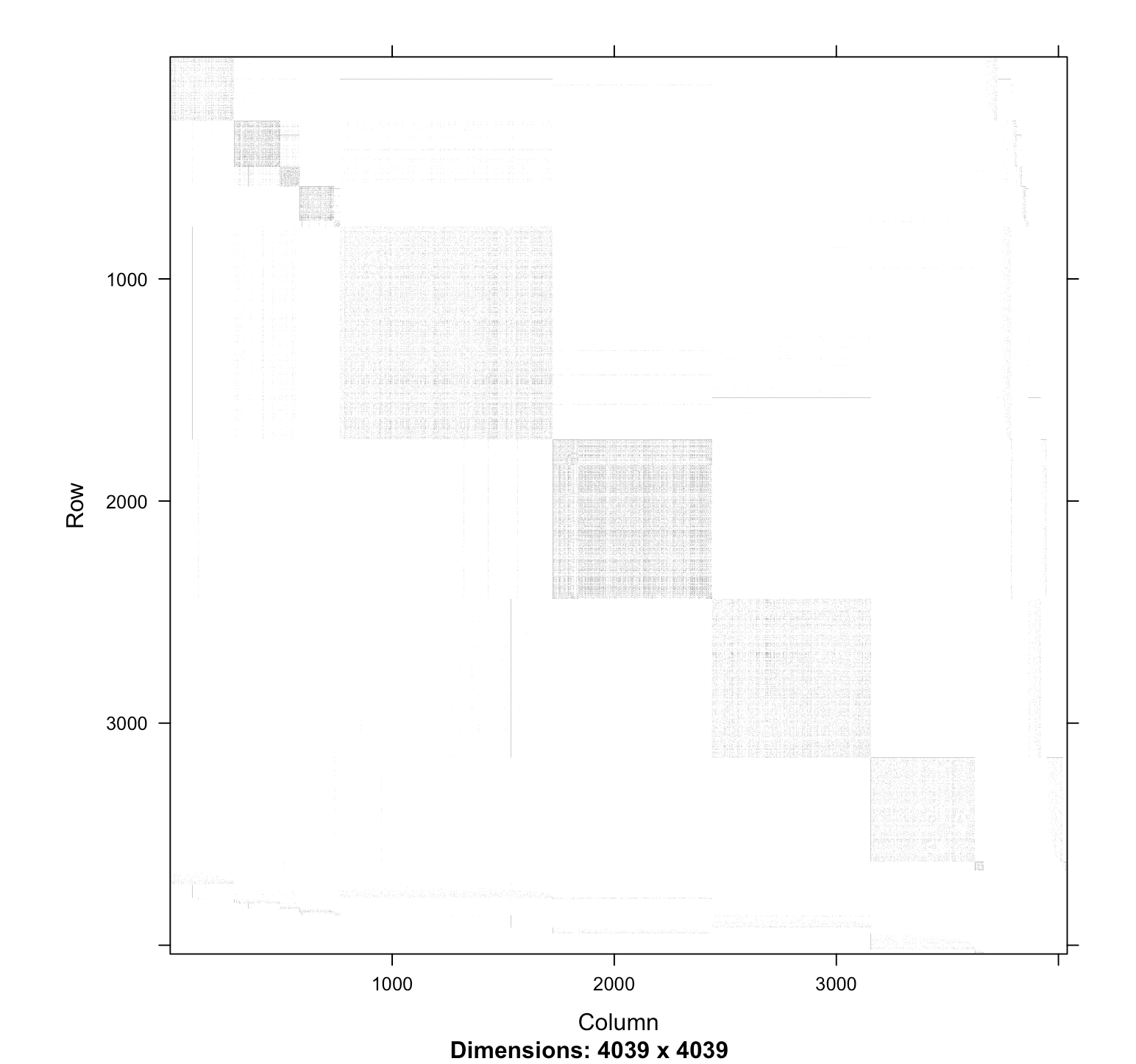} 
\caption{The adjacency matrix for the Facebook ego network; in the network, each node denotes a user and each link denotes two users are friends in Facebook. From the adjacency matrix, we can easily observe several groups of friend list.  }
\label{fig:egofb}
\end{figure}


\end{document}